\documentclass[12pt,a4paper]{article}

\usepackage[dvipsnames]{xcolor}
\usepackage{amsmath, amsthm, amssymb,bm,tikz}
\usepackage{physics}
\usepackage{graphicx}
\usepackage{placeins}
\usepackage{hyperref} 
\usepackage{cite}
\usepackage{dsfont}
\usepackage{comment}
\usepackage{cancel}
\usepackage[normalem]{ulem}

\title{Velocity of a Quantum Particle in a Classically Forbidden Region}
\author{
Christian Beck$^{1, \ast}$,~
Sheldon Goldstein$^2$,~
Dustin Lazarovici$^1$,\\
Roderich Tumulka$^3$,~and 
Nino Zangh\`{\i}$^4$ 
}
\date{June 30, 2026}

\begin{document}
\hyphenation{wave-guide wave-guides wave-num-ber wave-num-bers}
\newcommand{\x}[1]{{\color{blue}#1}}
\newcommand{\xx}[1]{{\color{ForestGreen}#1}}
\newcommand{\rt}[1]{{\color{Brown}#1}}
\newcommand{\ch}[1]{{\color{cyan}#1}}
\newcommand{\nin}[1]{{\color{magenta}#1}}

\newcommand{\be}{\begin{equation}}
\newcommand{\ee}{\end{equation}}
\renewcommand{\Re}{\mathrm{Re}}
\renewcommand{\Im}{\mathrm{Im}}
\newcommand{\Bu}{\text{{B\"u}}}
\newcommand{\dwell}{\mathrm{dwell}}
\newcommand{\EEE}{\mathbb{E}}
\newcommand{\PPP}{\mathbb{P}}

\maketitle

\noindent {}$^1$ Humanities and Arts Department, Technion---Israel Institute of
Technology, Haifa, Israel\\
{}$^2$ Departments of Mathematics and Physics, Rutgers University, Hill Center, 110 Frelinghuysen Road, Piscataway, NJ 08854-8019, USA\\
{}$^3$ Fachbereich Mathematik, Eberhard-Karls-Universit\"at T\"ubingen, Auf der Morgenstelle 10, 72076 T\"ubingen, Germany\\
{}$^4$ Dipartimento di Fisica, Universit\`a di Genova,
Via Dodecaneso 33, 16146 Genova, Italy
\&\ Istituto Nazionale di Fisica Nucleare (Sezione di Genova)
{}$^\ast$ Present address: Department of Philosophy, University of Graz, Heinrichstra\ss e 26/5, 8010 Graz, Austria
\bigskip

\begin{abstract}
Recently, Sharoglazova et al.\ [Nature 643, 67 (2025)] proposed a procedure for determining the speed of a quantum particle in the classically forbidden region of a potential step, and implemented it in a beautiful experiment.
The inferred speeds disagree significantly with the Bohmian velocities, which the authors presented as an experimental challenge to Bohmian mechanics. This is puzzling because the speeds are inferred from particle populations in coupled waveguides for which Bohmian mechanics and standard quantum mechanics make identical predictions. We resolve the puzzle by a detailed theoretical analysis of the experimental setup. We show that the speed inference rests on an assumption that fails in the relevant (evanescent) regime according to both Bohmian mechanics and standard quantum mechanics---namely that the inter-waveguide tunneling time is set by the transverse coupling and is not affected by entanglement with the longitudinal degree of freedom. We also consider a second speed estimate suggested by Sharoglazova et al., which is based on the B\"uttiker dwell time formula for particles in the forbidden region. We show that the authors applied the formula incorrectly, and that a correct application yields exact agreement with the predictions of Bohmian mechanics. Our analysis includes explicit calculations of Bohmian trajectories, dwell times, and longitudinal speeds in the two-dimensional waveguide model of the experiment.

\medskip

\noindent{\bf Key words:} Bohmian mechanics; dwell time; B\"uttiker formula; tunneling; potential step; double well.
\end{abstract}

\tableofcontents

\section{Introduction}
\label{sec:intro}

\begin{figure}[htbp]
  \centering
    \includegraphics[width = \textwidth]{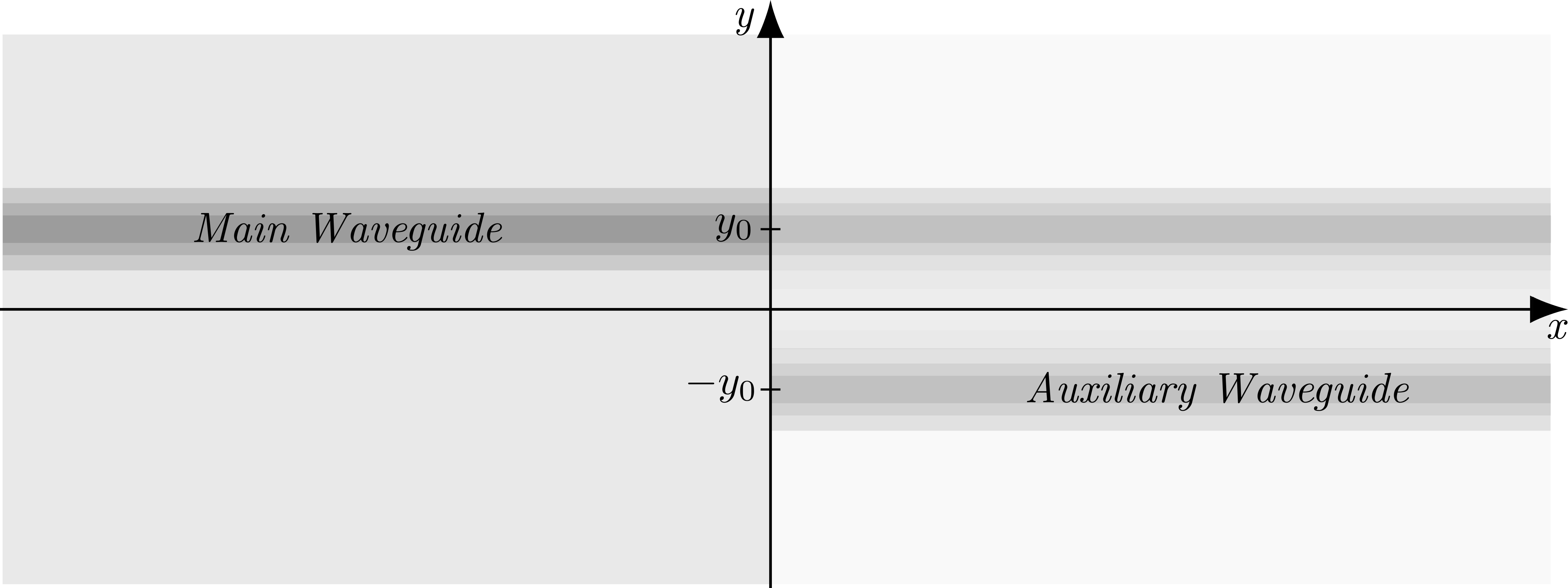}  
  \caption{Illustration of the potential landscape underlying the experimental setup (dark: low, light: high). The main waveguide at $y=y_0$ extends throughout; for $x \geq 0$ the transverse potential becomes a double-well, side-coupling the main waveguide to a parallel auxiliary waveguide at $y = -y_0$. At $x=0$, there is a potential step, acting as a barrier for the particle incident from the main waveguide. An explicit form is given in \eqref{2DV} below.}
\label{waveguides}
\end{figure}

When a quantum particle is tunneling through a barrier, or entering a potential step higher than its initial kinetic energy, it is moving in a classically forbidden region where its kinetic energy would be negative according to classical mechanics. In such a  region, the particle's wave function becomes \emph{evanescent}, characterized by an exponential spatial decay of its amplitude. 
Some authors have raised the question of what the relation between the speed of the particle and its kinetic energy would be when the kinetic energy appears to be negative. 
Sharoglazova et al.\ \cite{Sh25} recently approached this question by means of an innovative experiment aimed at measuring the speed of a quantum particle after it has entered a potential step. Their setup (using photons guided along coupled waveguides in a planar optical microcavity) effectively realizes the dynamics of a non-relativistic massive quantum particle moving in a two-dimensional potential landscape as depicted in Fig. \ref{waveguides}: For \(x<0\), the particle propagates along a main waveguide. At $x=0$, there is a potential step, and the main waveguide becomes side-coupled to a parallel auxiliary waveguide through a double-well structure in the transverse $y$-direction, allowing migration between the two waveguides.  The experimental parameters can be arranged such that the energy of the incoming particle lies below the height of the step (plus the ground-state energy of the transverse double well), thereby placing the longitudinal motion beyond the step in the classically forbidden (evanescent) regime. 

The key idea (first proposed in \cite{klaers.etal2023}) is that, since the coupling between the two waveguides is known, ``the respective populations of the two states can be understood as a measure of time'' \cite[p. 68]{Sh25}, which allows for the inference of an (undirected) $x$-speed in the classically forbidden region. The authors report that their estimated speed disagrees with the Bohmian velocity. At face value, this fact would seem to empirically refute Bohmian mechanics---or, at the very least, to pose, in the authors' own cautious words, a ``challenge'' to the Bohmian theory.

This is rather puzzling because the experiment would then seem to be able to empirically distinguish Bohmian mechanics from standard quantum mechanics, while it is well known that such a distinction is not possible~\cite{QE,QE2}. Even more basically, it is well known that the particle configuration in Bohmian mechanics has $|\Psi_t|^2$ distribution at every time $t$, and since the experiment only observes (the time average of) $|\Psi_t|^2$, it is clear that any finding based on this data cannot be in conflict with Bohmian mechanics. So how can the apparent contradiction be resolved? Which conclusions about particle motion in the classically forbidden region can legitimately be drawn from the experiment, and why? These are the questions we answer in this paper. Our discussion includes the computation of the Bohmian trajectories, average speed, and dwell time for the setup of the experiment. 

Sharoglazova et al.\ \cite{Sh25} claimed that the Bohmian velocity vanishes in the evanescent regime. This claim rests on the use of stationary scattering states, whereas (as we elucidate in Section~\ref{sec:stationary} and as pointed out in \cite{Dustin} and \cite{daem2026}) a precise analysis of the experiment requires time-dependent wave packets, for which the Bohmian velocity does not vanish. However---as also recently stressed by some of the original authors \cite{klaers2026}---this observation alone does not resolve the conflict because the Bohmian velocity induced by an evanescent wave packet is still
several orders of magnitude smaller than the speed reported by Sharoglazova et al. 

It turns out that the speed reported by Sharoglazova et al.\ is based on an assumption that sounds very plausible but upon scrutiny is found to be incorrect in both Bohmian and standard quantum mechanics; this is the ultimate resolution of this conflict. Let us explain the assumption. 

The double-well structure of the side-coupled waveguides in the $x>0$ region allows tunneling in the transverse $y$-direction between the two guides---a process distinct from the longitudinal tunneling at the potential step at $x=0$.  The basic idea behind the velocity estimate is that it is well known how much time it takes for a wave packet concentrated in the main waveguide to tunnel through the barrier separating the two guides and accumulate substantial weight in the auxiliary waveguide. When we observe substantial population (i.e., $|\Psi|^2$) in the auxiliary waveguide, we thus know that a certain time has elapsed. Given that substantial population in the auxiliary waveguide begins to build up at a certain longitudinal distance $X_{\rm eq}$, we can conclude that this distance is reached after a certain time, which allows us to estimate a corresponding longitudinal velocity. However, this reasoning rests on the assumption that the tunneling time in the $y$-direction is set by the transverse coupling between the waveguides and \emph{not influenced by the $x$-dependence of the wave function $\Psi$}. This is the assumption, analyzed in detail in Section \ref{sec:tunnelingtime}, that is not valid. 

Our analysis answers the challenge posed by Klaers et al.\ \cite{klaers2026}, to provide ``a non-circular argument showing why the speed inferred from the coupled waveguide system, which gives the expected result in the propagating regime, ceases to provide a meaningful speed measurement in the evanescent regime.'' We explain in detail why entanglement between the $x$ and $y$ degrees of freedom significantly affects the transverse tunneling dynamics in the evanescent regime but not in the propagating one.

The paper of Sharoglazova et al.\cite{Sh25} contains a second consideration based on the \emph{B\"uttiker formula} for the dwell time of particles in the forbidden region. It is presented by the authors as a consistency check on their speed estimate, but the same reasoning could be used as an independent speed estimate, yielding a value close to theirs but far from the Bohmian velocity. In Section~\ref{sec:dwell}, we point out that Sharoglazova et al.\ applied the dwell time formula incorrectly, and that a correct application yields a value that is far from theirs and in exact agreement with the Bohmian predictions.

\bigskip

The remainder of this paper is organized as follows. In Section~\ref{sec: model}, we summarize the setup and theoretical description of the experiment. In Section~\ref{sec:argument}, we present the two arguments for the speed estimates of Sharoglazova et al. In Sections~\ref{sec:stationary} and \ref{sec:1D}, we provide a theoretical analysis of the experiment (concerning the Schr\"odinger equation in Section~\ref{sec:stationary} and Bohm's equation of motion in Section~\ref{sec:1D}), yielding, for Bohmian mechanics, velocity estimates that are much smaller than those of Sharoglazova et al., thus leaving us with two puzzles. In Section~\ref{sec:tunnelingtime}, we explain the resolution of the first puzzle. In Section~\ref{sec:dwell}, we elucidate how to apply the B\"uttiker formula to the case at hand and thereby resolve the second puzzle. 
In Section~\ref{sec:conclusions}, we conclude. Further calculations are collected in Appendix~\ref{app:meanvel}.

\section{Setup}
\label{sec: model}

\begin{figure}[htbp]
  \centering
    \includegraphics[scale=0.7]{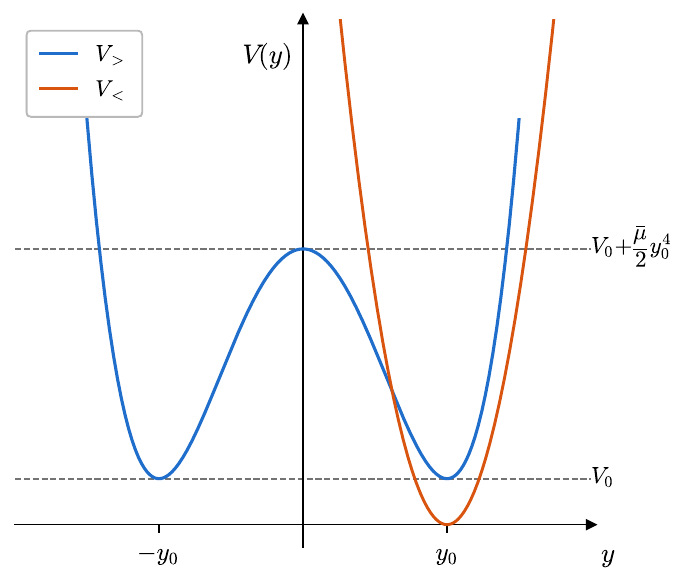}  
  \caption{Potentials $V_<$ and $V_>$ as in \eqref{2DV}}
\label{fig:potentials}
\end{figure}

We now provide equations for the potential $V(x,y)$ and solutions to the corresponding Schr\"odinger equation
\begin{equation}\label{eq:Schroedinger}
i\hbar\,\frac{\partial \Psi(x,y,t)}{\partial t}
= \left[ -\frac{\hbar^{2}}{2m}\!\left(\frac{\partial^{2}}{\partial x^{2}} + \frac{\partial^{2}}{\partial y^{2}}\right) + V(x,y) \right]\Psi(x,y,t).
\end{equation}
To model the experimental setup, we follow the suggestion of Sharoglazova et al. \cite{Sh25} that the coupling between the two waveguides corresponds to an effective double-well potential in the transverse direction. We make this concrete by considering a particle in a two-dimensional potential of the form
\begin{equation}\label{2DV}
V(x,y)=\begin{cases}
V_<(y)=2\mu y_0^2\bigl(y-y_0\bigr)^2 &\text{for }x<0 \\[2mm]
V_>(y)=V_0+\tfrac{1}{2}\bar\mu(y^2-y^2_0)^2 &\text{for }x\geq 0
\end{cases}
\end{equation}
with constants $\mu>0$, $\bar\mu >0$, $y_0>0,$ and $V_0>0$; the transverse cross sections of $V$, on either side of the step at $x=0$, are depicted in Figure~\ref{fig:potentials}. $V_<(y)$ is a harmonic potential with frequency $\omega = 2 y_0 \sqrt{\mu/m}$, providing the main waveguide along which the particle propagates for $x<0$.  $V_>(y)$ is a quartic double well with minima at $y= \pm y_0$; the well centered around $y_0$ continues the main waveguide, while the well around $-y_0$ forms the parallel auxiliary waveguide. 

It may appear natural to choose $\bar\mu=\mu$, so that the second derivative of $V_>(y)$ at the minima agrees with that of $V_<(y)$. However, we find it more convenient to choose them slightly differently so that the ground state energy of the harmonic oscillator $V_<$ agrees with the average energy $\bar{E}$, defined in \eqref{Eplusminus}, of the ground state doublet of the double-well potential  $V_{\rm dw}(y):= V_>(y) - V_0$, i.e.,
\be\label{omegaEbar}
\tfrac{1}{2}\hbar\omega = \bar{E}\,.
\ee
We assume the deep-well regime $y_0^{3}\sqrt{\bar\mu m}/\hbar \gg 1$,  so that the barrier height between the wells ($V_>(0) - V_>(\pm y_0)=\frac{1}{2}\bar\mu y_0^4$) greatly exceeds the harmonic zero-point energy ($\tfrac{1}{2} \hbar \bar\omega = \hbar y_0 \sqrt{\bar\mu/m}$) of each well.

\subsection{Double well}

We first collect some facts about the 1D  Schr\"odinger equation, only along the $y$-axis, of a double-well potential $V_{\rm dw}(y)= V_>(y) - V_0$, i.e., without the constant offset $V_0$, which will be accounted for separately:
\be\label{Schry}
i\hbar \frac{\partial \psi(y,t)}{\partial t}= \biggl[-\frac{\hbar^2}{2m} \frac{\partial^2}{\partial y^2}+V_{\rm dw} (y)\biggr]\psi(y,t)\,.
\ee

The analysis in the subsequent sections---like that of Sharoglazova et al.---does not depend on the exact form of the double-well potential but only on the well-known fact (see, e.g., \cite[Compl.~$\text{G}_{\text{IV}}$]{CohenTannoudji}\cite[\S 50]{LandauLifshitz}) that in the deep well regime, the two lowest energy eigenvalues are very close to each other, and the two corresponding eigenstates $\phi_+$ and $\phi_-$ can be written as symmetric and antisymmetric superpositions of states $\phi_m$ and $\phi_a$ (approximately) localized in one of the wells (provided that the phases of the eigenstates are chosen in the appropriate natural way). That is, 
\begin{equation}\label{pmmodes} \phi_{\pm} (y) = \frac{1}{\sqrt{2}} \left(\phi_m(y) \pm \phi_a(y) \right), 
\end{equation}
with corresponding energies $E_\pm$. (The subscripts refer to the main and auxiliary waveguides of the
two-dimensional model, cf.\ Sec.~\ref{sec:2Dstationary}.) We define\footnote{Note that the $\pm$ subscript labels the symmetric/antisymmetric state; hence $E_+ < E_-$. In terms of the potential parameters in \eqref{2DV}, the splitting $2\hbar J_0$ is exponentially small in the dimensionless parameter $y_0^3\sqrt{m \bar{\mu}} /\hbar$.}
\begin{equation}\label{Eplusminus} 
\bar{E} = \frac{E_++E_-}{2} \text{ and } J_0= \frac{E_--E_+}{2\hbar}, 
\end{equation}
so
\begin{equation}\label{Eplus}
E_{\pm} = \bar{E} \mp \hbar J_0\,. \end{equation}
The localized states $\phi_m$ and $\phi_a$ can be approximated by the harmonic-oscillator ground states corresponding to the quadratic approximation of the double-well potential around each minimum. The parameter $J_0$ (referred to as ``coupling constant'' in \cite{Sh25}) characterizes the energy splitting
$E_{-} - E_{+} = 2 \hbar J_0$
of the ground-state doublet. In this one-dimensional model, $J_0$ also sets the frequency of Rabi oscillations describing coherent tunneling between the two wells. These oscillations arise from coherent superpositions of the even/odd modes with their slightly different energies: 
\begin{equation}\begin{split} \label{1DdoubW}
&\frac{1}{\sqrt{2}} \left( e^{-i E_+ t/\hbar} \phi_+(y) +  e^{-i E_- t/\hbar} \phi_-(y) \right)\\
 =  &\frac{1}{\sqrt{2}} e^{-i \bar{E} t/\hbar} \left( e^{i J_0 t} \phi_+(y) +  e^{-i J_0 t} \phi_-(y) \right)\\
 = & e^{-i \bar{E} t/\hbar} \Bigl( \cos(J_0 t) \phi_m(y) +  i \sin(J_0 t) \phi_a(y) \Bigr).
\end{split}\end{equation}

\subsection{Scattering states for a step potential}
\label{sec:step_problem}

We next collect some facts about the other 1D  Schr\"odinger equation, only along the $x$-axis, 
\be\label{Schrx}
i\hbar \frac{\partial \psi(x,t)}{\partial t}= \biggl[-\frac{\hbar^2}{2m} \frac{\partial^2}{\partial x^2}+V(x) \biggr]\psi(x,t)\,,
\ee
with a step potential
\be
V(x) = V_0\,\Theta(x), \qquad V_0 > 0,
\ee
where \(\Theta(x)\) is the Heaviside step function, equal to \(0\) for \(x<0\) and \(1\) for \(x\ge 0\).

\paragraph{Propagative regime}
If the kinetic energy of an incident plane wave,
\be
E_k = \frac{\hbar^2 k^2}{2m},
\ee
satisfies \(E_k > V_0\), the stationary Schr\"odinger equation admits scattering eigenfunctions
\be\label{propscatteringstates}
\varphi_k(x) =
\begin{cases}
e^{ikx} + \mathrm{r}(k)\, e^{-ikx}, & x < 0, \\[1mm]
\mathrm{t}(k)\, e^{i\bar{k}(k) x}, & x > 0,
\end{cases}
\ee
with
\be\label{kbar}
\mathrm{r}(k) = \frac{k - \bar{k}}{k + \bar{k}}, 
\qquad
\mathrm{t}(k) = \frac{2k}{k + \bar{k}},
\qquad
\bar{k}(k) = \frac{1}{\hbar}\sqrt{{2m}(E_k - V_0)}.
\ee

\paragraph{Evanescent regime.}
If \(E_k < V_0\), the incident plane wave has energy below the step, formally corresponding to an imaginary wavenumber $\bar{k}(k) = i\kappa(k)$ in the classically forbidden region. The stationary scattering states thus take the form
\begin{equation}\label{varphi_k}
\varphi_k(x) =
\begin{cases}
e^{ikx} + e^{2i\theta(k)}\, e^{-ikx}, & x < 0, \\
\gamma(k) e^{i\theta(k)}\, e^{-\kappa(k) x}, & x > 0,
\end{cases}
\end{equation}

\noindent with
\be\label{rtkappasub}
e^{2i\theta(k)} = \frac{ik + \kappa(k)}{ik - \kappa(k)}, 
\quad
\kappa(k) = \sqrt{k_c^2-k^2}, \quad
\gamma(k) = \frac{2k}{k_c}, \quad k_c= \frac{\sqrt{2mV_0}}{\hbar}.
\ee
It will be convenient to multiply \eqref{varphi_k} by the phase factor $e^{-i\theta(k)}$, leading to the real-valued states
\begin{equation}\label{realscatteringstates}
\tilde{\varphi}_k(x) =
\begin{cases}
e^{-i\theta(k)}e^{ikx}+e^{i\theta(k)}e^{-ikx}=  2\cos\!\left(kx - \theta(k)\right), & x < 0, \\[4pt]
\gamma(k) e^{-\kappa(k)x}, & x > 0.
\end{cases}
\end{equation}
In any case, inside the potential step, the propagating wave is replaced by an exponentially decaying evanescent tail with decay constant $\kappa(k) >0$.

\subsection{Two-dimensional double well model}
\label{sec:2Dstationary}

Now we combine the longitudinal scattering with the double-well model for the transverse coupling between the waveguides. Since the two lowest eigenvalues of the double well are close to each other whereas the next eigenvalue is not close, it is reasonable to assume (as we will henceforth and as Sharoglazova et al.\ did) that none of the higher excited states of the double well gets populated; that is, we use a two-level approximation in the region $x\geq 0$. Likewise, for $x<0$, we assume that only the ground state for the harmonic potential $V_<$ gets populated. This ground state, in turn, can be approximated by $\phi_m$, the state localized in the corresponding well of $V_>$ that extends the main waveguide. Therefore, the relevant Schr\"odinger equation becomes
\be
i\hbar \frac{\partial \Psi(x,y,t)}{\partial t} = H\Psi(x,y,t),
\ee
with
\be
H=-\frac{\hbar^2}{2m}  \frac{\partial^2}{\partial x^2}+\begin{cases}
\frac{1}{2}\hbar\omega |\phi_m\rangle \langle \phi_m| &\text{for }x<0\\[2mm]
V_0 + E_+ |\phi_+\rangle \langle \phi_+| + E_-|\phi_- \rangle \langle \phi_-| &\text{for }x\geq 0
\end{cases}
\ee
and energies $E_\pm$ as in \eqref{Eplus}.

\paragraph{Propagative regime.}
Incoming longitudinal waves exciting the even and odd double well modes acquire, instead of $\bar k$  as in \eqref{kbar}, corresponding wavenumbers
\begin{equation}\label{wavenumbers}k_\pm(k) = \frac{1}{\hbar} \sqrt{2m \left(E_k- V_0  \pm \hbar J_0\right)},
\end{equation}
respectively, where $E_k = \frac{\hbar^2 k^2}{2m}$ is the longitudinal kinetic energy of the incoming wave. The ground state energy of the harmonic confinement for $x<0$ is by construction equal to the average energy $\bar{E}$ of the ground state doublet for $x>0$, see \eqref{omegaEbar},
so that the two contributions cancel in the energy-conservation balance, leaving 
\begin{equation}\label{energy-balance}
E_k = \frac{\hbar^2 k_\pm^2}{2m} + V_0 \mp \hbar J_0.
\end{equation}

Since $\phi_m = \tfrac{1}{\sqrt{2}}(\phi_+ + \phi_-)$ by \eqref{pmmodes}, a wave incident in the main waveguide excites an equal-weight superposition of the even and odd double well modes. In the propagative regime\footnote{Sharoglazova et al.~\cite{Sh25} define $\Delta := E - V_0 + \hbar J_0$, where their $E$ is shifted by an overall constant from our $E_k$; $\Delta$ equals $E_k - V_0$ in our notation, and the propagative/evanescent boundaries $E_k - V_0 = \pm \hbar J_0$ correspond to $\Delta = \pm \hbar J_0$.} ($E_k > V_0 + \hbar J_0$), a corresponding stationary scattering state in the step region $x>0$ thus takes the form
\begin{equation}\label{psikprop0}
\psi_k(x, y) = \frac{1}{\sqrt{2}} \left( \mathrm{t}_+(k) \, e^{i k_+ x} \phi_+(y) +  \mathrm{t}_-(k) e^{i k_- x} \phi_-(y) \right).
\end{equation}
We introduce the combinations
\be 
\delta k :=\frac{k_+ -\, k_-}{2} >0, \qquad \bar{k} :=\frac{k_+ +\, k_-}{2}.
\ee

\begin{figure}[htbp]
  \centering
     \includegraphics[width=\textwidth]{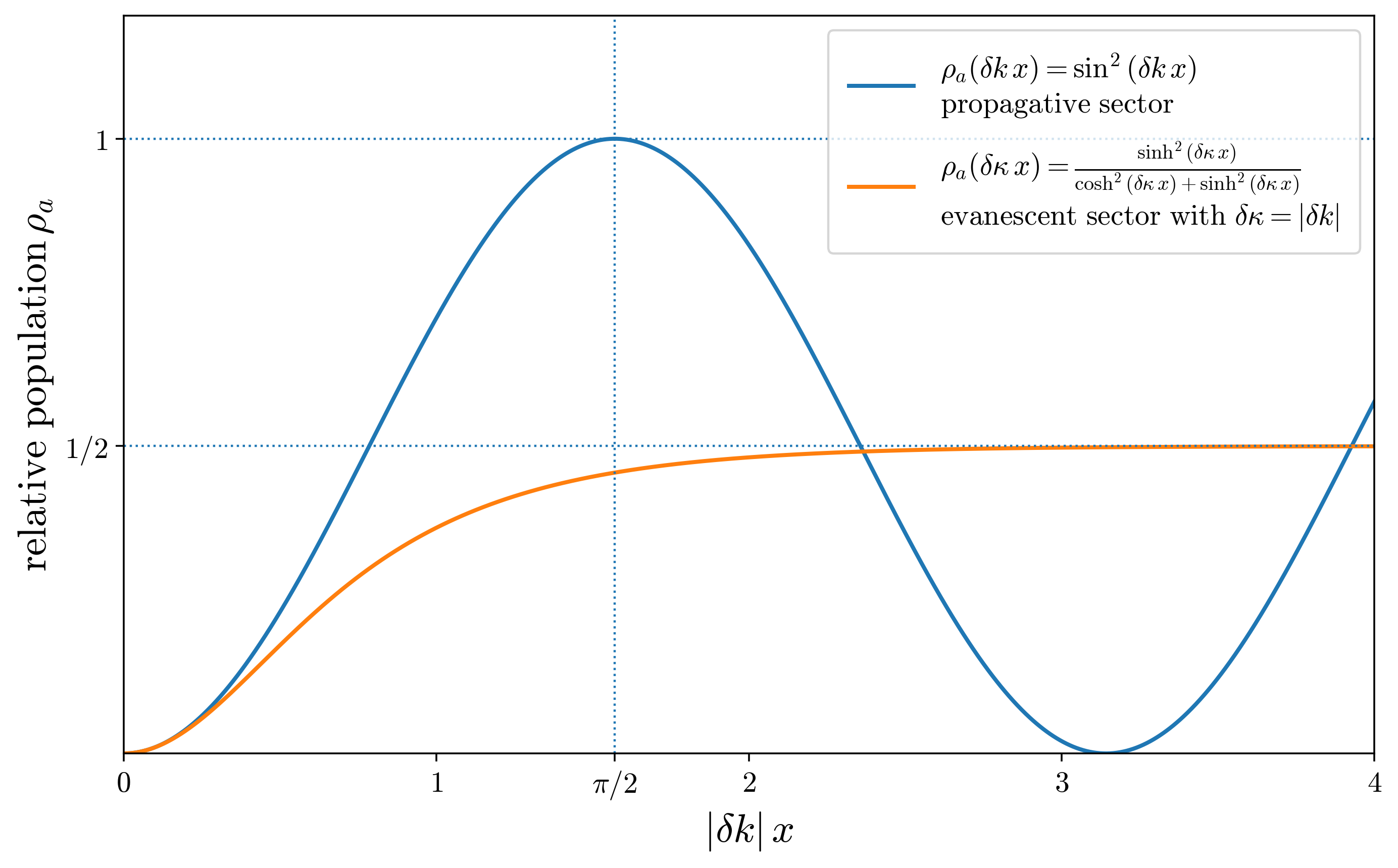}  
 \caption{Relative population $\rho_a$ in the auxiliary waveguide as a function of longitudinal distance from the step in units of $|\delta k|^{-1}$.  In the propagative regime (blue curve), the populations oscillate between the two waveguides, with complete transfer ($\rho_a = 1$) at odd multiples of $\pi/(2 \delta k)$. In the evanescent regime (orange curve, as given by \eqref{rhoa}), the populations equilibrate monotonically, approaching $\rho_a = 1/2$ over the characteristic length scale $X_{\mathrm{eq}} \sim 1/\delta\kappa$.}
\label{Fig:relativepopulation}
\end{figure}

For a small splitting $\delta k \ll k$, we can approximate both transmission amplitudes by the corresponding expression for the mean wavenumber, i.e.,
\be\label{tmean}  \mathrm{t}_\pm(k)\approx  \mathrm{t}(k):= \frac{2k}{k+\bar{k}}. \ee 
Equation \eqref{psikprop0} then simplifies to
\begin{equation}\begin{split}\label{psikprop}
\psi_k(x, y) & \approx \frac{\mathrm{t}(k)}{\sqrt{2}} \left( e^{i k_+ x} \phi_+(y) +  e^{i k_- x} \phi_-(y) \right)\\
& = \frac{\mathrm{t}(k)}{\sqrt{2}} \, e^{i\bar{k} x} \left( e^{i \delta k \, x} \phi_+(y) + e^{- i \delta k \, x} \phi_-(y) \right)  \\
& = \mathrm{t}(k) \, e^{i \bar{k}x } \, \bigl( \cos(\delta k \, x) \phi_m(y) +  i \sin(\delta k \, x) \phi_a(y) \bigr).
\end{split}\end{equation}
The final equality in \eqref{psikprop} uses the relation \eqref{pmmodes} between the even/odd modes \(\phi_\pm\) and the localized transverse modes \(\phi_m,\phi_a\). In the localized basis, one readily sees that the state exhibits spatial beating between the main and auxiliary waveguides, with the first complete population transfer from the main to the auxiliary waveguide occurring at $x=\frac{\pi}{2\delta k}$ (blue curve in Fig. \ref{Fig:relativepopulation}).

\paragraph{Evanescent regime.}
In the evanescent regime ($E_k < V_0 -\hbar J_0$), the wavenumbers \eqref{wavenumbers} become imaginary, $k_{\pm} = i \kappa_{\pm}, \, \kappa_{\pm} \in \mathbb{R}^+$. Note that $\kappa_+ < \kappa_-$. Starting from \eqref{realscatteringstates}, the stationary states in the classically forbidden region $x>0$ take the form 
\begin{equation}\begin{split}\label{realexponentials}
    \psi_k(x, y) & = \frac{1}{\sqrt{2}} \, \Bigl(\gamma_+ e^{- \kappa_+ x} \phi_+(y) +  \gamma_- e^{- \kappa_- x} \phi_-(y) \Bigr)\\
    & \approx \frac{\gamma}{\sqrt{2}} \, \Bigl(e^{- \kappa_+ x} \phi_+(y) +  e^{- \kappa_- x} \phi_-(y) \Bigr)\\
    & = \gamma e^{- \bar{\kappa} x} \, \Bigl( \cosh(\delta\kappa \, x) \, \phi_m(y) +  \sinh(\delta\kappa \, x) \, \phi_a(y) \Bigr),
\end{split}\end{equation}
with 
\be \label{kappa12}
\delta\kappa :=\frac{\kappa_- -\, \kappa_+}{2} >0, \qquad \bar{\kappa} :=\frac{\kappa_- +\, \kappa_+}{2},
\ee
and 
\be\label{meangamma} \gamma_\pm(k)\approx \gamma(k):= \frac{2k}{\sqrt{k^2+\bar{\kappa}^2(k)}}. 
\ee
The evanescent regime therefore does not produce oscillatory population transfer. Instead, the different decay constants $\kappa_\pm$ filter the even and odd components unequally, with the antisymmetric mode attenuated more strongly than the symmetric one: 
\be \frac{e^{-\kappa_-x}}{e^{-\kappa_+x}}
= e^{-(\kappa_- - \kappa_+)x} = e^{-2\delta\kappa \, x} \to 0, \text{ as } x\to \infty\,.
\ee
The result is a redistribution of weight from the main to the auxiliary waveguide, with the relative population in the auxiliary guide, 
\be\label{rhoa} \rho_a(x) := \frac{\sinh^2(\delta\kappa \, x)}{\sinh^2(\delta\kappa \, x) + \cosh^2(\delta\kappa \, x)}, \ee
approaching $1/2$ over the characteristic equilibration length\footnote{Throughout the paper we use the following notational convention: $\approx$ denotes approximate equality, $\sim$ same order of magnitude, $\propto$ equality up to a multiplicative constant.}
\begin{equation}\label{Xeq}
X_{\rm eq}\sim  \frac{1}{\delta\kappa},
\end{equation}
see Fig. \ref{Fig:relativepopulation}.
We omit discussion of the mixed regime ($V_0 - \hbar J_0 \leq E_k \leq V_0 +\hbar J_0$), where the incident energy lies above barrier for the symmetric ground state of the transverse double well but not for the slightly higher-energetic anti-symmetric mode.

\section{The Main Arguments}
\label{sec:argument}

Building on the waveguide model of the previous section, we now present two arguments for estimating, in the evanescent regime, the speed of quantum particles in the classically forbidden region---our best reconstruction of the reasoning found in \cite{Sh25} and \cite{klaers.etal2023}. The first is based on the tunneling time between the two waveguides; the second on a well-known formula for the dwell time of particles inside the potential step.

\subsection{Tunneling time} \label{sec:tunneltimearg}

The tunneling time argument goes as follows. Since velocity is distance over time, to infer a speed from an observed population transfer at a longitudinal distance of order $X_{\rm eq} \sim {1}/{\delta\kappa}$, one needs the corresponding time. Using the tunneling time scale $T_{\rm tun}\sim {1}/{J_0}$ between the two wells of the transverse double-well potential (from \eqref{1DdoubW}), we  obtain an average longitudinal velocity 
\be \label{vnaive}
v_x \sim \frac{X_{\rm eq}}{T_{\rm tun}}\sim\frac{J_0}{{\delta\kappa}}
\ee  
for a particle in the classically forbidden region.\footnote{Note that the key relation $\delta\kappa\, x = \tfrac{J_0}{v_x}\, x$ employed by Sharoglazova et al.\ \cite{Sh25} is equivalent to \eqref{vnaive}. Since $J_0$ has been determined independently, only $\delta\kappa$ remains to be measured. In the reported experiment, this was done by fitting the small-$x$ behavior of the relative auxiliary-waveguide population to $\rho_a(x) \approx (\delta\kappa\, x)^2$---the leading-order expansion of \eqref{rhoa}.}
Using the relation
\be 
{\delta\kappa} \, \bar{\kappa}=\frac{mJ_0}{\hbar},
\ee 
equation \eqref{vnaive} becomes
\be \label{longvelK2}
v_x\sim\frac{\hbar\bar{\kappa}}{m}.
\ee 

Notably, the analogous argument gives the uncontroversially correct result in the propagative regime. With population transfer between the waveguides achieved at the longitudinal distance $X_{\rm tun} =\frac{\pi}{2\delta k}$ and the corresponding tunneling time $T_{\rm tun}= \frac{\pi}{2J_0}$, their ratio
\be\label{vprop}
v_x = \frac{J_0}{\delta k} = \frac{\hbar \bar{k}}{m}
\ee
is precisely the group velocity inside the potential step, as represented by the factor $e^{i\bar{k}x}$ in \eqref{psikprop}. 

(Together with \eqref{longvelK2} this suggests a simple relationship between particle speeds and---what might appear to be---the  kinetic energy,  $E_\mathrm{kin}=\frac{\hbar^2 k^2}{2m}-V_0$:
\be\label{vEkin}
v=\sqrt{2|E_\mathrm{kin}|/m},
\ee
valid for both positive and negative $E_\mathrm{kin}$.)

\subsection{Dwell time} \label{sec:dwelltime}

As mentioned in Section~\ref{sec:intro}, an observation presented by Sharoglazova et al.\ \cite{Sh25} as a check on their speed estimate \eqref{longvelK2} can actually be regarded as an independent method for determining the velocity of particles in the evanescent region. We  call this method the dwell time argument.
For this argument, it suffices to consider only the 1D  step potential in the $x$-direction (see Section \ref{sec:step_problem}). (Since ${\delta\kappa} \ll \bar{\kappa}$, what is relevant for the dwell time argument in 2D  differs little from that for 1D.) 

The dwell time argument goes as follows. Consider a one dimensional scattering problem with incident plane wave $ e^{ikx}$ where $k>0$, and the corresponding stationary scattering state $\varphi_k$ of mode $k$.  A widely accepted expression for the average dwell time of the particle in a region $(a,b)$ is given by the B\"uttiker formula \cite{Bu83} 
\be \label{dwell1}
   \tau_d(k) = \frac{m}{\hbar k} \int_a^b \mathrm{d}x \, |\varphi_k(x)|^{2}=(j_\mathrm{in})^{-1}\int_a^b \mathrm{d}x \, |\varphi_k(x)|^{2}
\ee 
with the incident current $j_\mathrm{in}=\frac{\hbar k}{m}$ (see Section \ref{sec:dwell} for derivation and discussion). 

For a step potential $V(x)=V_0\,\Theta(x)$, the stationary evanescent states take the form \eqref{varphi_k}, i.e., $\varphi_{k}(x)=\frac{2ik}{ik-\kappa}e^{-\kappa x}$ for $x>0$.  With $(a,b) = (0, \infty)$, Eq. \eqref{dwell1} thus gives the dwell time in the classically forbidden region
\be \label{dwell2}
 \tau_d(k) =\frac{4k^2}{k^2+\kappa^2} \left. \left(-\frac{1}{2\kappa}e^{-2\kappa x}\right)\right|_{x=0}^{\infty}\frac{m}{\hbar k} =\frac{2m}{\hbar\kappa}\frac{k}{k^2+\kappa^2}.
\ee
Estimating the typical penetration depth by the decay length $\lambda=\frac{1}{2\kappa}$ of the evanescent mode, this dwell time \eqref{dwell2} corresponds to a velocity
\be \label{vdwell}
v_x\sim\frac{2\lambda}{\tau_d} =\frac{\hbar\kappa}{m}\left(\frac{1+\beta^2}{2\beta}\right),
\ee 
with $\beta(k):= k / \kappa$. This completes the dwell time argument. 

The estimate \eqref{vdwell} can be compared with the previous, tunneling-time-based estimate \eqref{longvelK2}. In particular, the factor $\left(\frac{1+\beta^2}{2\beta}\right)$ in \eqref{vdwell} is of order $1$ as long as the kinetic energy $\frac{\hbar^2 k^2}{2m}$ of the incoming particle is not too close to zero and not too close to $V_0$. In this energy regime, \eqref{vdwell} is thus of the same order of magnitude as \eqref{longvelK2}. 
This is more or less the argument given by Sharoglazova et al.\ \cite{Sh25} (except that they introduce \eqref{vdwell} not as an independent estimate but present the good agreement of \eqref{vdwell} and \eqref{longvelK2}---over an energy range where $\beta \approx 1$---as corroborating the latter).

In the following sections, we will compare the velocity formulas obtained from the tunneling time and dwell time arguments with the predictions of Bohmian mechanics and see that they indeed conflict. However, both arguments rely on implicit assumptions whose validity is not guaranteed. In Sections \ref{sec:tunnelingtime} and \ref{sec:dwell}, we discuss these assumptions, explain why they fail in the evanescent regime, and show that, once corrected, no discrepancy remains.

\section{Stationary States versus Wave Packets}
\label{sec:stationary}

Sharoglazova et al.\ \cite{Sh25} suggested not merely a quantitative mismatch between their estimated speed and Bohmian velocities, but that Bohmian particles do not move at all in the evanescent region. The claim is based on the fact that for a stationary evanescent state \eqref{realexponentials}, the probability current---and hence the Bohmian velocity---vanishes identically. As already pointed out in \cite{Dustin}, this would make it puzzling from the perspective of both Bohmian mechanics and standard quantum mechanics how any particle population could build up there. The basic mistake here is taking stationary scattering theory too literally as a physical description.

Stationary scattering theory provides a powerful description of scattering processes, but its stationary eigenfunctions are \emph{not} physical states of a single particle at a given time.  
They are non-normalizable monochromatic waves of infinite extent, representing idealized components of a scattering process.  
As such, they encode the \emph{modulus} and \emph{phase} of each spectral component, but they do not describe the actual transient evolution of a localized particle (cf. \cite{NLMK09}). 

A real scattering event involves a \emph{wave packet} with finite spectral width. The physical dynamics---in both standard quantum mechanics and Bohmian mechanics---emerge from the coherent superposition of spectral components and their time evolution. 
For the speed estimates sketched in the previous section, stationary scattering theory---understood as capturing the characteristics of the dominant spectral component of a narrow-band wave packet---was sufficient. For the  more detailed analysis of Sections \ref{sec:1D}--\ref{sec:dwell}, it won't be sufficient and we need to consider the time-dependent dynamics of a narrow-band wave packet penetrating the step region.

\subsection{Spectral representation and narrow--band approximation}
\label{sec:narrowband}

Again, we start with the 1D evanescent case, for the $x$-direction only. 
We assume that the incoming spectrum \(\widehat{\psi}_{\mathrm{in}}(k)\) has a {\it narrow-band} Gaussian profile, i.e., one that is narrowly peaked around some \(k_0\), $0<k_0<k_c$:

\be\label{Gaussian}
\widehat{\psi}_{\mathrm{in}}(k)=\frac{1}{(2\pi\sigma_k^2)^{1/4}}\,
\exp\!\left[-\frac{(k-k_0)^2}{4\sigma_k^2}\right],
\ee 
with $\sigma_k \ll \min\lbrace k_0, k_c-k_0\rbrace$ and $k_c$ as in \eqref{rtkappasub}. The resulting wave packet 
\be\label{psiint2}
\psi(x,t) =\frac{1}{\sqrt{2\pi}} \int_0^{k_c} \widehat{\psi}_{\mathrm{in}}(k)\, \tilde{\varphi}_k(x) \, e^{-i \omega(k) t} \, \mathrm{d}k,
\ee
built up from the real-valued scattering states $\tilde{\varphi}_k$ of \eqref{realscatteringstates} with $E_k = \hbar \omega(k)=\tfrac{\hbar^2 k^2}{2m}$, satisfies 
\begin{equation}\label{timereversal}
    \psi(x,-t) = \psi^*(x,t),
\end{equation}
i.e., time-reversal symmetry. This reflects the fact that the wave packet transiently penetrates the classically forbidden region, reaches its turning point at $t=0$ (in our coordinates), and is ultimately fully reflected. (Note that $\psi(x,t)$ is  almost but not quite normalized since $\widehat{\psi}_{\mathrm{in}}(k)$ is normalized on $(-\infty, \infty)$ rather than $(0,k_c)$---a discrepancy that we can and shall ignore).

Focusing on the step region $x>0$, we have  

\be\label{psiintx0}
\psi(x,t) = \frac{1}{\sqrt{2\pi}}\int_0^{k_c} \widehat{\psi}_{\mathrm{in}}(k)\,  \gamma(k)\,e^{-\kappa(k)x} e^{-i \omega(k) t} \, \mathrm{d}k.
\ee

For a first-order spectral approximation (valid while the $x$-dependent exponential does not shift the dominant spectral weight far from $k_0$), we linearize \(\kappa(k)\) and \(\omega(k)\) around the dominant wavenumber \(k_0\):
\be
\kappa(k)\approx \kappa_0-\beta_0(k-k_0),
\qquad
\omega(k)\approx \omega_0+\omega_0'(k-k_0),
\ee
where
\be
\kappa_0=\kappa(k_0),\qquad
\omega_0=\frac{\hbar k_0^2}{2m},\qquad
\omega_0'=\frac{\hbar k_0}{m},
\qquad
\beta_0=\frac{k_0}{\kappa_0}.
\ee
Freezing the factor $\gamma(k)$ at $\gamma_0 := \gamma(k_0)$  and extending the range of integration to $k \in (-\infty, +\infty)$, one obtains

\begin{equation}\label{psi_compact}
\psi(x,t)\approx 
\gamma_0\left(\frac{2 \sigma_k^2}{\pi}\right)^{1/4}
\,e^{-\kappa_0 x}\,e^{-i\omega_0 t}
\exp\!\left[
\sigma_k^2 \,
\Big(\beta_0 x -i \omega_0' t\Big)^2
\right].
\end{equation}

At $t=0$, the exponent becomes purely real-valued, and---for a narrow-band packet---we have for the density [for $0<x\ll 1/(\sigma_k\beta_0)$] that
\be\label{psi2trev2}
\rho(x):= |\psi(x,0)|^2 \approx   \rho (0) e^{-2\kappa_0 x}, \quad \rho(0)= \gamma_0^2\,\sigma_k\sqrt{2/\pi}. 
\ee

This expression captures the exponential decay associated with the dominant spectral component at \(k_0\), and will be sufficient for the integrated quantities considered below. Note, however, that for any fixed $t$, the function \eqref{psi_compact} is actually not square-integrable in $x$ due to the positive $x^2$-term in the argument of the exponential. This underscores that the approximation is valid only for $x$ that is not too large.\footnote{\label{fn:erfc} Maintaining the exact upper limit of integration, \(k\le k_c\), in \eqref{psiintx0} while extending the lower limit to \(-\infty\) yields the more complicated but square-integrable approximation
$$\psi(x,t)\approx
\frac{\gamma_0}{2} \left(\frac{2\sigma_k^2}{\pi}\right)^{{1}/{4}} e^{-\kappa_0 x-i\omega_0 t}
\exp\!\left[\sigma_k^2(\beta_0 x-i\omega'_0 t)^2\right]
\operatorname{erfc}\!\left(\frac{k_0-k_c+2\sigma_k^2(\beta_0 x-i\omega'_0 t)}{2\sigma_k}\right).$$ However, recovering the correct behavior of the large-$x$ tails---which is not controlled by the dominant spectral peak, but by the threshold region \(k\to k_c\), where \(\kappa(k)\to 0\)--- requires a different analysis that will be presented elsewhere. For present purposes, the approximation \eqref{psi_compact} is sufficient.}

\subsection{Two-dimensional wave packet}
The two-dimensional evanescent wave packet 
\be \label{psi2D0}
\Psi(x,y,t) = \frac1{\sqrt{2\pi}}\int\limits_0^{k^\downarrow_c}\widehat{\psi}_\mathrm{in}(k) \, \psi_k(x,y) e^{-i\omega(k)t} \mathrm{d}k
\ee is built up from the scattering states \eqref{realexponentials} with $k< k^\downarrow_c :=\hbar^{-1}\sqrt{2m(V_0-\hbar J_0)}$, the critical wavenumber for exciting the symmetric ground state of the transverse double well. The scattering states  $\psi_k$ are superpositions of longitudinal evanescent waves coupled to the transverse double-well modes $\phi_\pm$: \begin{equation}\begin{split}\label{psi2Dint}
\Psi(x,y,t) &= \frac{1}{\sqrt{2\pi}} \sum_{s=\pm} \frac{\phi_s(y)}{\sqrt{2}} \int\limits_0^{k^\downarrow_c} \widehat{\psi}_\mathrm{in}(k) \, \gamma_s(k) e^{-\kappa_s(k)x} e^{-i\omega(k)t} \mathrm{d}k \\
&\approx \frac{1}{\sqrt{2\pi}} \sum_{s=\pm} \frac{\phi_s(y)}{\sqrt{2}} \int\limits_0^{k^\downarrow_c} \widehat{\psi}_\mathrm{in}(k) \, \gamma(k) e^{-\kappa_s(k)x} e^{-i\omega(k)t} \mathrm{d}k
\end{split}\end{equation}
for $x>0$, where the approximation arises again from assuming $\lvert \kappa_+ - \kappa_-\rvert \ll k_0$ and replacing both factors $\gamma_\pm(k)$ by the corresponding expression for the mean decay constant \eqref{meangamma}.

Using the respective narrow-band approximation \eqref{psi_compact} for each of the two summands in \eqref{psi2Dint}, we get 

\be\label{psi2D_shited} \Psi(x,y,t) \approx \gamma_0\left(\frac{\sigma_k^2}{2\pi}\right)^{1/4} e^{-i\omega_0t} \sum_{s=\pm} {\phi_s(y)} e^{-\kappa_s x}\, \exp\!\left[\sigma_k^2 \,
	\Big(\beta_s x-i\omega_0't\Big)^2\right], \ee
with $\beta_\pm = {k_0}/{\kappa_\pm}$. Since the transverse modes $\phi_\pm(y)$ are orthogonal, the marginal density
\be
\rho_t(x) = \int  \lvert \Psi(x,y,t) \rvert^2 \, \mathrm{d}y
\ee
is just the arithmetic mean of the one-dimensional densities $|\Psi_\pm(x,t)|^2$ associated with each mode. In particular, in the limit $\sigma_k \to 0$, we have at $t=0$,
\begin{equation}\label{rhoy}
\rho(x) \propto \frac{1}{2} \left(e^{-2\kappa_+ x} + e^{-2\kappa_- x}\right)
\end{equation}
in the sense discussed above.

\section{Bohmian Predictions}
\label{sec:1D}

In this section, we will analyze Bohmian particle motion in the classically forbidden region. As explained above, capturing the Bohmian velocity in the evanescent region requires a treatment in terms of time-dependent wave packets (of finite spectral width) rather than stationary scattering states. 

When a wave packet with energy below the barrier impinges on the step, it is ultimately reflected with probability 1, yet a transient evanescent wave develops in the classically forbidden region.  In ordinary quantum mechanics, this wave produces a small but finite current that reverses and ultimately vanishes as the packet is reflected. In Bohmian mechanics, the same current produces a slow forward motion of trajectories in the forbidden region, followed by a reversal at a well-defined turning time (here, $t=0$), after which they retrace their paths and exit the forbidden region. Nonetheless, as we will see, there is a significant discrepancy between the Bohmian velocity in the classically forbidden region and the velocity estimates given in Section \ref{sec:argument} following Sharoglazova et al. 

Since the essential physics is already captured by the one-dimensional problem of a sub-threshold wave packet scattering off a step potential, we shall focus on this simpler model first, before discussing how the results extend to the two-dimensional model with coupled waveguides.

\subsection{Bohmian velocity in the forbidden region (1D)}\label{sec:forbidden_field}

The Bohmian velocity field is
\be
v(x,t)=\frac{J(x,t)}{\rho(x,t)},
\ee
with $J=\frac{\hbar}{m}\,\Im(\psi^*\partial_x\psi)$ and $\rho=|\psi|^2$. Equivalently, writing $\psi$ in polar form as
\be
\psi=R\exp(iS/\hbar)
\ee
with real $R$ and $S$,
\be
v(x,t) = \frac{1}{m}\partial_x S.
\ee
From \eqref{psi_compact}, we can see that in the classically forbidden region 
\be
    S(x,t) = -\hbar\omega_0 t - 2\hbar\sigma_k^2
\,\beta_0 x \, \omega_0't,
\ee
so
\be\label{valphat}
v(x,t) = - \alpha t, 
\ee
with
\be\label{alphadef}
\alpha := \frac{\hbar}{m}2\sigma_k^2 \beta_0\omega_0' = \frac{2\hbar^2 \sigma_k^2 k_0^2}{m^2\kappa_0} \,.
\ee
Three observations: $v$ is independent of $x$, $v$ is a linear function of $t$, and the sign of $v$ changes from plus to minus at $t=0$. This means that the different Bohmian trajectories (in the region $x>0$) are translates of each other, that they are parabolas
\be\label{Xt1D}
X(t)=X(0)-\frac{\alpha}{2}t^2\,,
\ee
and that they move to the right for $t<0$, turn around at the reversal time $t=0$, and move to the left for positive times $t>0$. Each trajectory enters the forbidden region, penetrates to a maximal depth $\xi = X(0)$, and returns (cf. Fig. \ref{fig:2Dtraj}b below).

\subsection{Bohmian average dwell time and speed (1D)}

Let $\xi=X(0)$ denote the maximal depth of a Bohmian trajectory. As a consequence of \eqref{Xt1D}, the time required to travel the distance $\xi$ is 
\be \tau(\xi) = \sqrt{\frac{2 \xi}{\alpha}}.\ee
The distribution of the turning points is given by $\rho(\xi) = |\psi(\xi,0)|^2$. In the narrow-band regime, we have from \eqref{psi2trev2} that 
\be\label{psi2trev}
\rho(x) \approx \rho(0) e^{-2\kappa_0 x}.
\ee
The average dwell time of trajectories entering the evanescent region is thus 
\begin{align}
   \tau_\mathrm{B} &=2 \,\mathbb{E}(\tau(\xi) \mid \xi>0 )  = 2 \frac{\int_0^\infty \tau(\xi) \, \rho(\xi)\,  \dd \xi}{\int_0^\infty  \rho(\xi)\,  \dd \xi}\\
   &\approx  2 \frac{\int_0^\infty \sqrt{2\xi/\alpha} \, e^{-2\kappa_0\xi}\,  \dd \xi}{\int_0^\infty  e^{-2\kappa_0\xi}\,  \dd \xi} = \sqrt{\frac{\pi}{\alpha \kappa_0}} = \sqrt{\frac{\pi}{2}} \frac{m}{\hbar k_0 \sigma_k},\label{sojourn}
\end{align}
where the factor $2$ accounts for the return journey after the turnaround. This is in stark disagreement with \eqref{dwell2}.

\bigskip

We can also compute the average speed $\bar{v}_\mathrm{B}$ of a Bohmian particle in the forbidden region $x>0$. For an individual trajectory turning around at $x=\xi >0$, the mean speed in the forbidden region is
\be
v(\xi)= \frac{\xi}{\tau(\xi)}=\sqrt{\frac{\xi\alpha}{2}}. 
\ee

The ensemble average---conditional on trajectories spending a positive amount of time in the forbidden region---is thus

\begin{equation}\begin{split}\label{vB}
\bar{v}_B &= \mathbb{E}\left( v(\xi) \mid \xi>0\right) =  \frac{\int_0^\infty v(\xi) \, \rho(\xi)\,  \dd \xi}{\int_0^\infty  \rho(\xi)\,  \dd \xi}\\
   &\approx \sqrt{\frac{\alpha}{2}}  \frac{\int_0^\infty \sqrt{\xi} \, e^{-2\kappa_0\xi}\,  \dd \xi}{\int_0^\infty  e^{-2\kappa_0\xi}\,  \dd \xi} =  \sqrt{\frac{\pi\alpha}{16\kappa_0}}
= \sqrt{\frac{\pi}{2}} \frac{\hbar  k_0 \sigma_k}{2m\kappa_0}.
\end{split}\end{equation}

We observe that this speed is proportional to $\sigma_k$, the spectral width of the wave packet, and thus approaches $0$ in the monochromatic limit $\sigma_k \to 0$. In particular, \eqref{vB} is very different from the speed estimates \eqref{longvelK2} or \eqref{vdwell} based on the reasoning of Sharoglazova et al.---in fact, orders of magnitude slower for the parameters relevant to their experiment.\footnote{\label{footnote6}The experiment uses a pulse duration of $26$\,ns (full width at half maximum, FWHM). Treating this as the temporal width of the wave packet gives $\sigma_k \approx 10^{-5}\,\mu\mathrm{m}^{-1}$. For the deeply evanescent run reported in \cite{Sh25} ($k_0 \approx 0.30\,\mu\mathrm{m}^{-1}$, $\kappa_0 \approx 0.12\,\mu\mathrm{m}^{-1}$), this yields $\bar{v}_B \approx 0.24\,\mathrm{km/s}$, compared to $\hbar\kappa_0/m \approx 1800\,\mathrm{km/s}$.} Furthermore, in contrast to the relation \eqref{vEkin} suggested by Sharoglazova et al., the Bohmian velocity $\bar{v}_B$ \eqref{vB} depends on $\sigma_k$ and  not just on the value of $E_{\mathrm{kin}}=\hbar^2 k_0^2/2m-V_0$.

\subsection{Bohmian trajectories for the 2D  waveguide model}
\label{sec:2Ddetailed}
In the relevant approximation, the two-dimensional analysis mirrors the one-dimensional case in key respects. For the evanescent wave packet \eqref{psi2D_shited}, two observations make this precise. 

First, since for $\bar{\kappa} \gg \delta\kappa$, i.e., $\kappa_+ \approx \kappa_-$ the phases of the two terms on the right hand side of \eqref{psi2D_shited} are almost the same, and since the set of configurations for which they interfere destructively have negligible probability,  we have for the longitudinal Bohmian velocity that

\begin{equation}\label{meanvel}
v_x^\Psi(x,y,t)
\approx 
-\alpha t,
\qquad
\alpha:=
\frac{2\hbar^2\sigma_k^2k_0^2}{m^2\bar{\kappa}}.
\end{equation}
This is the same relation as \eqref{valphat} and \eqref{alphadef} in the 1-dimensional case, except that the decay constant $\kappa_0$ gets replaced by the average decay constant $\bar\kappa$. In particular, the $x$-motion is here independent of the $y$-coordinate.

Second, as in the one-dimensional case, the wave function satisfies
\begin{equation}\label{timereversal2}
\Psi(x,y,-t)= \Psi^*(x,y,t),
\end{equation}
i.e., time-reversal symmetry. The Bohmian velocity field is therefore odd in $t$,
\begin{equation}
v^\Psi(x,y,-t) = - v^\Psi(x,y,t),
\end{equation}
so both velocity components (longitudinal and transverse) reverse simultaneously at $t=0$, and each trajectory turns around and retraces exactly the path by which it entered the classically forbidden region. 

\begin{figure}[ht]
\centering
  \includegraphics[width=\textwidth]{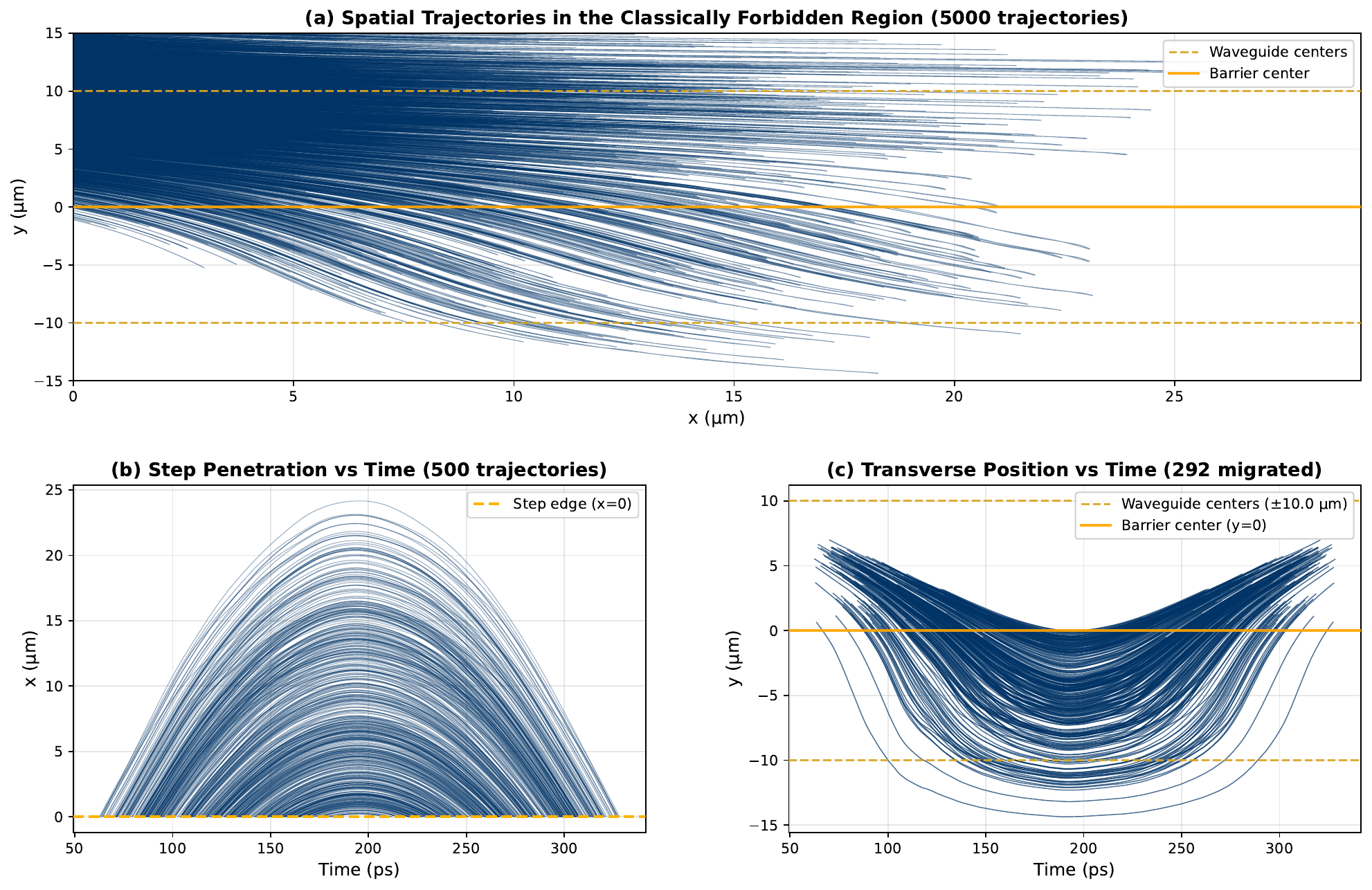}
  \caption{\label{fig:2Dtraj} Numerical simulation of Bohmian trajectories in the classically forbidden region ($x>0$) of the 2D setup for $V_0=0.54\, \mathrm{meV}, \hbar J_0\, = 26.2 \,\mu\mathrm{eV}, E_{k_0}-V_0 = -0.07\,\mathrm{meV},\sigma_k = 2.5\times 10^{-3}\,\mu\mathrm{m}^{-1}$. (a) Spatial paths of 5,000 trajectories sampled from $|\Psi|^2$. Main waveguide centered around $y_0 = 10\,\mu\mathrm{m}$; auxiliary waveguide centered around $-y_0 = -10\,\mu\mathrm{m}$.  (b) Longitudinal coordinate $X(t)$ for 500 representative trajectories, exhibiting parabolic turnaround. Horizontal axis centered around median reversal time. (c) Transverse coordinate $Y(t)$ for 292 trajectories that crossed $y=0$ prior to turnaround; the simultaneous reversal of $X(t)$ and $Y(t)$ reflects the time-reversal symmetry \eqref{timereversal2}.}
\end{figure}

This behavior is confirmed numerically in Fig.~\ref{fig:2Dtraj}, where we also see a subset of trajectories migrating from the main into the auxiliary waveguide prior to turnaround.

\subsection{Bohmian average speed in 2D}

Since the $x$-motion of the Bohmian particles in the 2D  case is the same as in the 1D  case except for $\bar\kappa$ replacing $\kappa_0$, also the average $x$-speed in the classically forbidden region is obtained from \eqref{vB} through this replacement,
i.e.,
\be\label{vB2}
\bar{v}_\mathrm{B} =\sqrt{\frac{\pi}{2}} \frac{\hbar \sigma_k k_0}{2m\bar{\kappa}}.
\ee

In contrast, the tunneling time argument estimates an average speed over the characteristic equilibration distance $X_{\rm eq} \sim \frac{1}{\delta\kappa}$, where the populations in the main and auxiliary waveguides become roughly equal.  Since $X_{\rm eq}$ is significantly greater than the decay length $\lambda= \frac{1}{2\bar{\kappa}}$, only a small fraction of particles---with substantially above-average speeds---reach such depths in the forbidden region.  For a more appropriate comparison, we should thus consider, instead of \eqref{vB}, the average Bohmian speed $\bar{v}'_{\rm B}$ in the forbidden region conditional on trajectories reaching $X_{\rm eq}$, that is,
\begin{equation}\label{vBprime}
\bar{v}'_{\rm B} = \mathbb{E}\left( v(\xi) \mid \xi\geq X_{\rm eq}\right) =  \frac{\int_{X_{\rm eq}}^\infty v(\xi) \, \rho(\xi)\,  \dd \xi}{\int_{X_{\rm eq}}^\infty  \rho(\xi)\,  \dd \xi},\end{equation}
with $\rho(x) \propto\frac{1}{2}\left(e^{-2\kappa_+ x} + e^{-2\kappa_- x}\right)$ as per \eqref{rhoy}. Since $X_{\rm eq} \gg 1/\kappa_+ \approx 1/\kappa_-$, contributions from \(\xi \gg X_{\rm eq}\) are exponentially suppressed, so the  average is dominated by the contribution from $\xi$ near $X_{\rm eq}$, i.e.,\footnote{In a more precise analysis, the integrals can be evaluated in closed form:
$$
\bar{v}'_{\rm B} = 
\sqrt{\frac{\alpha}{2}}\,
\frac{\sum_{\pm}(2\kappa_\pm)^{-3/2}\Gamma\!\left(\frac{3}{2},2\kappa_\pm X_{\rm eq}\right)}
{\sum_{\pm}(2\kappa_\pm)^{-1}e^{-2\kappa_\pm X_{\rm eq}}},
$$
where $\Gamma(s,t)$ is the upper incomplete gamma function with $\Gamma(\tfrac{3}{2}, z) = \frac{1}{2}\sqrt{\pi}\operatorname{erfc}(\sqrt{z}) + \sqrt{z}e^{-z}$.
In the regime $2\bar{\kappa}X_\mathrm{eq}=\bar{\kappa}/\delta\kappa\gg 1$, we can use the expansion $\Gamma\left(\frac{3}{2}, z\right) = \sqrt{z}e^{-z}\left(1+\mathcal{O}(z^{-1})\right)$, which leads to $
\bar{v}'_{\rm B} = \sqrt{\frac{\alpha X_\mathrm{eq}}{2}} \left[1 + \mathcal{O}\left(\frac{\delta\kappa}{\bar{\kappa}}\right)\right].$}
\begin{equation}\label{vXeq}
\bar{v}'_B \approx v(X_{\rm eq}) = \sqrt{\frac{\alpha X_\mathrm{eq}}{2}} = \frac{\hbar}{m}\frac{\sigma_k\,k_0}{\sqrt{\bar{\kappa}\,\delta\kappa}} \,.
\end{equation}
This exceeds $\bar{v}_\mathrm{B} $ by a factor 
\begin{equation}
\frac{\bar{v}'_\mathrm{B}}{\bar{v}_\mathrm{B}} \propto \sqrt{\frac{\bar{\kappa}}{\delta\kappa}}, 
\end{equation}
reflecting the fact that $\bar{v}_\mathrm{B} $ is dominated by trajectories reaching only the decay length $\lambda= \frac{1}{2\bar{\kappa}}$. Nonetheless, \eqref{vXeq} still differs significantly from the speed $\sim\hbar \bar{\kappa}/{m}$ proposed by Sharoglazova et al.

We thus have the following puzzle. Bohmian mechanics is empirically equivalent to standard quantum mechanics and predicts, in particular, the same Born distributions for particle populations in the waveguides---which is the primary data observed in the experiment. At the same time, Bohmian mechanics describes particle speeds in the evanescent region that are (though not zero as claimed in \cite{Sh25}) orders of magnitude smaller than the estimates obtained from either the tunneling time or the dwell time argument sketched in Section  \ref{sec:argument}. So, how can this fit together? The following sections will provide the resolution.

\FloatBarrier

\section{Resolution: Tunneling Time in the Double Well and Entanglement}
\label{sec:tunnelingtime}

As discussed in Section \ref{sec:argument}, the speed inferred by Sharoglazova et al.\ relies on the assumption of a tunneling time $T_{\rm tun} \sim \frac{1}{J_0}$ for migrations between the waveguides. This assumption, in turn, is motivated by the fact that $J_0$ sets the frequency of oscillations between the localized states in the double-well model, see \eqref{1DdoubW}. 
While prima facie plausible, this reasoning overlooks the fact that, as we are going to explain, \emph{entanglement} between longitudinal ($x$) and transverse ($y$) degrees of freedom can have a significant effect on the tunneling dynamics in the double well. It turns out that, owing to the entanglement, the correctness of the naive tunneling time $T_{\rm tun} \sim \frac{1}{J_0}$, set by the waveguide coupling, cannot be taken for granted, and in the evanescent regime, it is, in fact, false. 

Clearly, one could not assume the naive tunneling time if there were a significant \emph{dynamical} coupling between the longitudinal and transverse degrees of freedom, i.e., if the total Hamiltonian contained an interaction term $H_{\rm int}(x,y)$ that couples the longitudinal motion directly to transitions between transverse modes. Such a term would, in general, prevent the transverse dynamics from being describable by an autonomous two-level Hamiltonian and thus invalidate any tunneling time inferred from the coupling $J_0$ alone. While the present setup does involve an explicit 
$x$-dependence of the transverse potential \eqref{2DV},
\be
V(x,y) = \mathds{1}_{x < 0}(x)V_{<}(y) + \mathds{1}_{x\geq 0}(x)V_{>}(y),
\ee
this is \emph{not} the issue here. Indeed, the setup could be modified by extending the double-well structure (i.e., the coupling between the waveguides) to $x<0$ and adjusting the source--step distance so that the incident wave packet is concentrated in the main waveguide near $x=0$. Such a modification would remove the $x$-dependent switching of the transverse potential without significantly affecting the relevant physics.

However, even in the absence of an explicit interaction term, entanglement between the $x$- and $y$-degrees of freedom can disrupt the tunneling dynamics in the transverse double well. To illustrate this phenomenon and isolate the relevant mechanism from the specific details of the waveguide geometry, the following minimal model is instructive: Instead of one particle in a two-dimensional setup, consider a system of two particles, each with one degree of freedom. The motion of the first particle (coordinate $x \in [0, L)$) is confined to a ring ($\cong S^1$) of circumference $L$. The kinematics of the second particle (coordinate $y$) correspond to the double-well model. The total Hamiltonian takes the separable form 
\be
H = H_{x,0} + H_y,
\ee
where $H_{x,0}$ is the free Hamiltonian on the ring and $H_y$ includes the double-well potential $V_{\rm dw}(y)=\tfrac{\bar{\mu}}{2}(y^2-y^2_0)^2$. Let $\phi_\pm(y)$ denote the even and odd eigenstates of the double well with energies $E_{\pm} = \bar{E} \mp \hbar J_0$, and let $\chi_\pm(x):= \sqrt{\frac{2}{L}}\sin(k_{\pm} x)$ be real standing-wave eigenfunctions of $H_{x,0}$ with 
\be k_\pm = \sqrt{\frac{2m}{\hbar^2} \left(E  - E_{\pm}\right)} = \frac{2\pi n_\pm}{L} \ee  for an energy $E> E_-$ and for appropriately chosen $L$.\footnote{The consistency condition, ensuring that $\chi_\pm$ satisfy periodic boundary conditions, is $\frac{\hbar^2}{2m}\left(\frac{2\pi}{L}\right)^2 (n_+^2-n_-^2) = E_- - E_+ = 2\hbar J_0, \; n_{+}>n_{-} \in \mathbb{N}$. Note that the lower-energy double-well mode $\phi_+$ pairs with the higher kinetic-energy ring mode $\chi_+$
(and vice versa), since both branches must compensate to the same total energy.} Now consider the entangled state 
\begin{equation}\label{es}
\Psi(x,y,t) = e^{-iEt/\hbar}\frac{1}{\sqrt{2}}\left(\chi_+(x)\phi_+(y) + \chi_-(x)\phi_-(y)\right).
\end{equation}
This is a stationary state of the composite system of energy $E$ with none of the ``pathologies'' of the scattering states \eqref{realexponentials}, i.e., it is normalized and not evanescent. It involves no interactions between the $x$ and the $y$ particles; however, because the $\chi_\pm$ are orthogonal, the two particles are maximally entangled. Their Bohmian velocity field vanishes identically, since the wave function can be chosen real up to a global phase. In terms of the localized double-well modes $\phi_m$ and $\phi_a$, \eqref{es} becomes
\begin{equation}\label{es1}
\Psi(x,y,t) = e^{-iEt/\hbar}\sqrt{\frac{2}{L}}\bigl(\cos(\delta k\, x)\sin(\bar{k}x)\phi_m(y) + \sin(\delta k \, x)\cos(\bar{k}x)\phi_a(y)\bigr),
\end{equation}
where $\delta k= \frac{k_+-k_-}{2}$ and $\bar{k}= \frac{k_++k_-}{2}$. Thus $\phi_m$ dominates the amplitude near $x= \pm \pi/\delta k, \pm 2\pi/\delta k, \ldots$ (and $\phi_a$ dominates near $x = \pm \pi/(2\delta k), \pm 3\pi/(2\delta k), \ldots$).  But this is a completely stationary profile. Even though $\phi_m$ and $\phi_a$ are coupled at the level of the Hamiltonian---via an off-diagonal matrix element $\propto J_0$---there are no oscillations and no tunneling between the wells, neither in terms of the Schr\"odinger dynamics, nor in terms of Bohmian particle motion. From an observed population ratio $\rho_a(x)$ in the wells, no corresponding time can be inferred at all.

This illustrates that entanglement can have a radical effect on the double well dynamics, so that the assumption of the naive tunneling time---on which the speed estimate \eqref{longvelK2} is based---is not justified, in general.

\begin{figure}[htpb]
  \centering
  \includegraphics[width=0.8\textwidth]{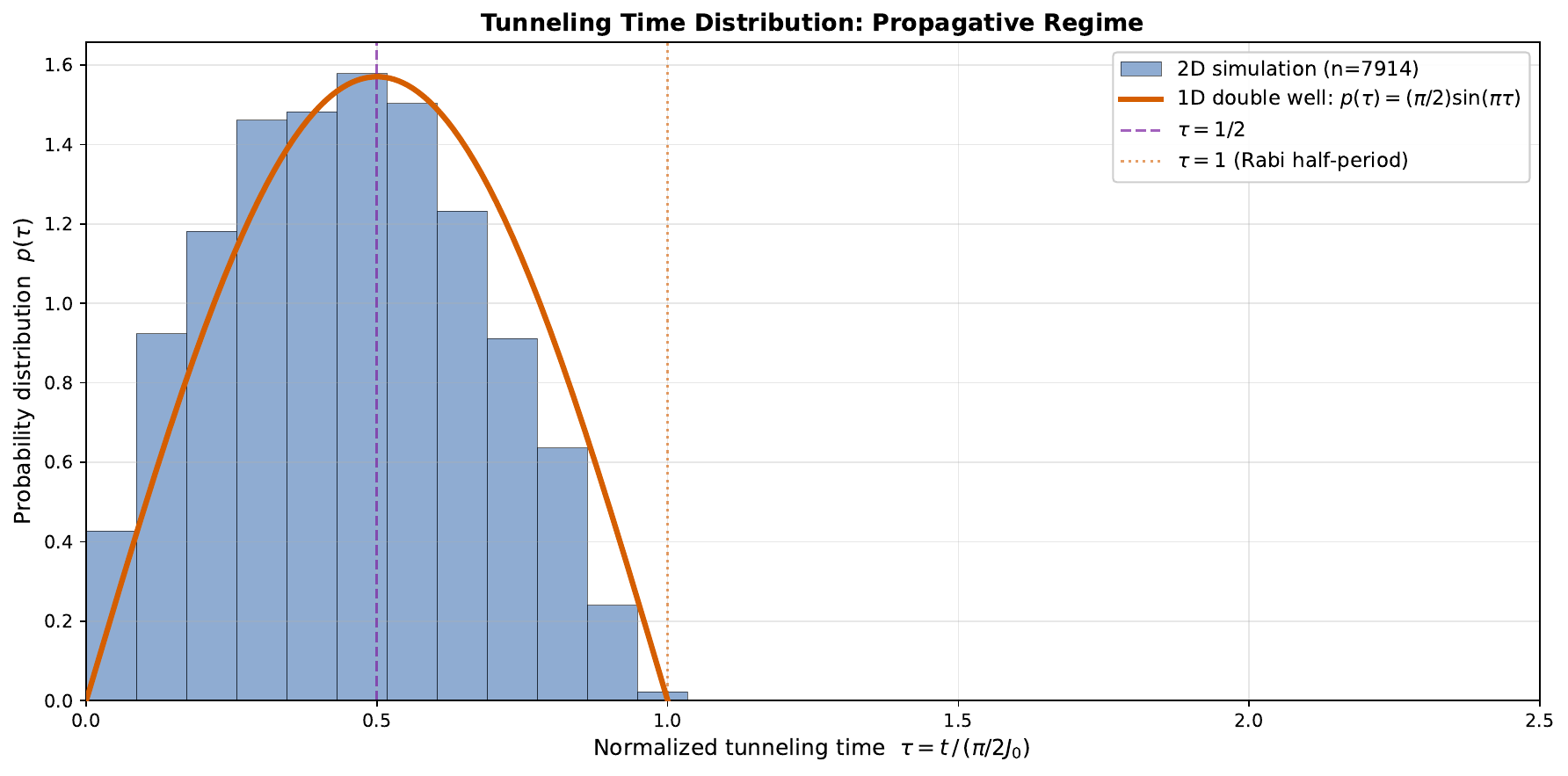}
   \includegraphics[width=0.8\textwidth]{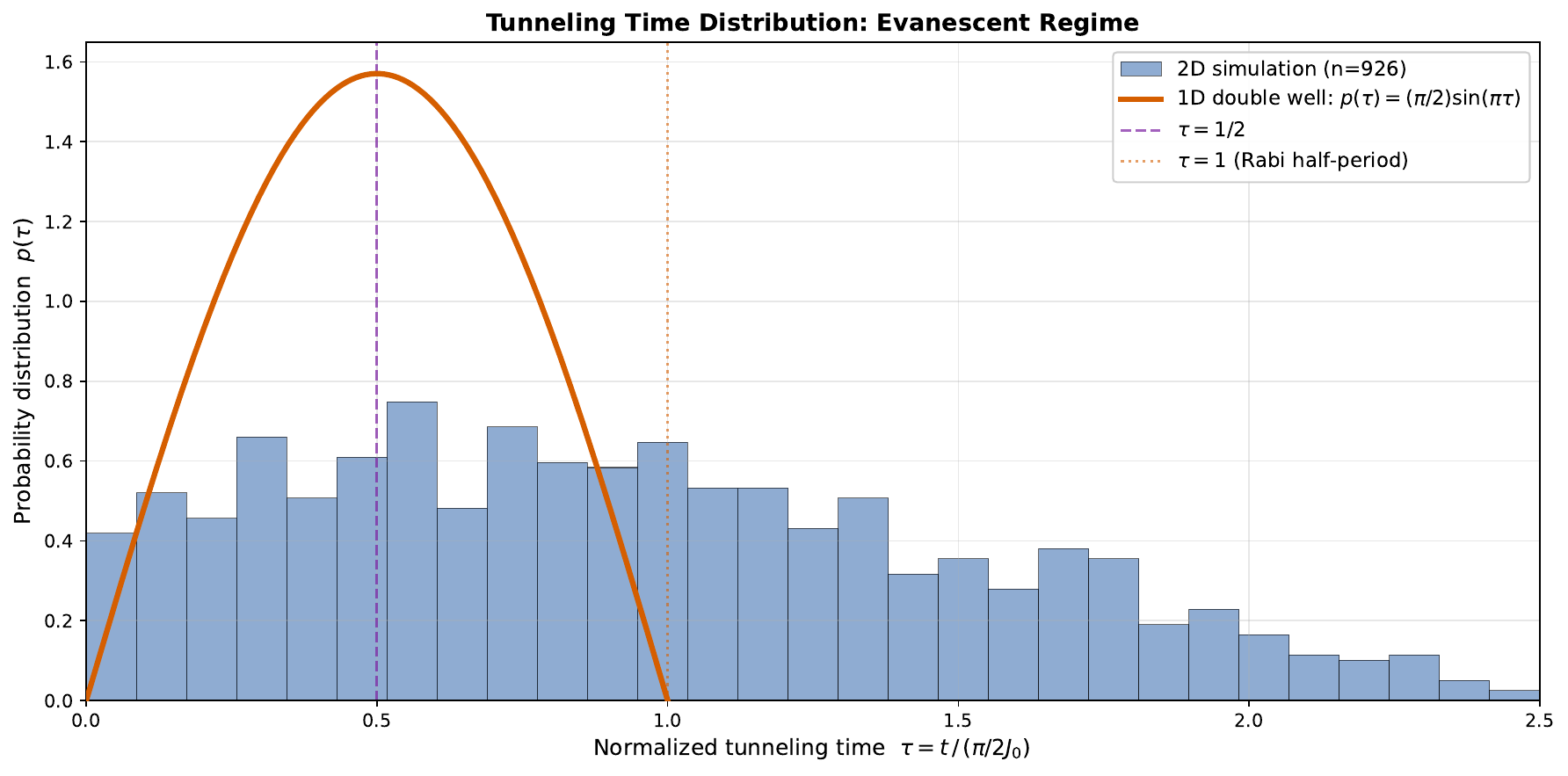}
  \caption{Distribution of tunneling times for Bohmian trajectories migrating from the main into the auxiliary waveguide of the 2D setup. Here, ``tunneling time'' is defined as the time the Bohmian particle takes from entering the classically forbidden region at $x=0$ (in the main waveguide, $y>0$) until entering the auxiliary waveguide, $y<0$. Top: Propagative regime ($E_{k_0} - V_0 = +0.15 \,\mathrm{meV}$). Bottom: Evanescent regime ($ E_{k_0} - V_0 = -0.07\,\mathrm{meV}$). $\sigma_k = 2.5\times 10^{-3}\,\mu\mathrm{m}^{-1}$ for both. Time is shown in units of $\pi/(2J_0)$. The orange curve shows for comparison the  distribution of the first time of crossing from $y>0$ to $y<0$, $p(\tau) \approx (\pi/2)\sin(\pi\tau)$, in the 1D double well model \eqref{1DdoubW} with initial state $\phi_m$.}
  \label{fig:BohmMigrationTime}
\end{figure}

For the experiment of Sharoglazova et al., Figure~\ref{fig:BohmMigrationTime} shows that this unjustified assumption---that the transverse tunneling time is unaffected by the $x$-dependence of $\Psi(x,y)$---is, in fact, false in Bohmian mechanics. It shows, first in the propagative and then in the evanescent case, the statistical distribution (in the ensemble of Bohmian trajectories for the same wave function) of the transverse tunneling time $T_\mathrm{tun}$, defined here as the time from entering the step region $x>0$ until the first crossing of $y=0$ in the transverse direction (if such crossing occurs). The analogous distribution for the 1D double well \eqref{1DdoubW} is shown in orange. The key observation is that the two histograms are not the same; specifically, the distribution in the 2D propagative case agrees with the 1D model, while in the evanescent case it does not. (The $y=0$ threshold is permissive, since trajectories entering the step near the symmetry plane can cross it without genuinely transitioning to the auxiliary waveguide, producing an excess at short tunneling times.)

\subsection{When does the naive tunneling time apply?}

This raises the question as to when the naive tunneling time \emph{does} apply. It certainly applies for a pure, separable double-well state---one that is neither dynamically coupled to nor entangled with other degrees of freedom and hence undergoes Rabi-style oscillations with frequency $J_0$. However, realistically speaking, the assumption that the double-well degree of freedom is perfectly separable---i.e., that the total state remains an exact product between $y$ and all other degrees of freedom---is hardly ever justified. 

On the other hand, any entangled state can be decomposed into a superposition of product states,
\begin{equation} \label{decompintoprod}
    \Psi_0(x, y) = \sum\limits_{i=1}^{n}  \psi_i(x)\phi_i(y),
\end{equation}
where the $\psi_i\neq 0$ are mutually orthogonal (but not in general normalized) and $n\in \mathbb{N}\cup\lbrace \infty\rbrace$. 
Under such a decomposition of the total wave function, the $y$-system has a well-defined state $\phi_i$ relative to each $\psi_i$ (given by the partial inner product $\phi_i \propto \langle \psi_i \lvert \Psi_{0}\rangle_x$). 
Assuming a separable Hamiltonian $H = H_x + H_y$ (more precisely, $H = H_x\otimes I + I\otimes H_y$), the unitary Schr\"odinger evolution factorizes, so that 
\begin{equation}\label{factorization}
    \Psi(t) = e^{-\frac{i}{\hbar} H t}\Psi_0 = \sum\limits_{i=1}^n \left(e^{-\frac{i}{\hbar} H_x t}\psi_i\right)\otimes\left(e^{-\frac{i}{\hbar} H_y t}\phi_i\right). 
\end{equation}
Hence, with the $\psi_i(t)$ evolving unitarily under $H_x$ (which preserves their mutual orthogonality), the corresponding relative states $\phi_i(t) = \langle \psi_i(t)\lvert \Psi(t) \rangle_x$ evolve autonomously under $H_y$. The question thus becomes: under what conditions can we think of the $y$-system as being effectively described by one of those relative states, that is, by   $\phi_1(y,t)$ \emph{or} $\phi_2(y, t)$ \emph{or} \ldots. Notably, since such a decomposition into relative states is always available, orthogonality of the $\{\psi_i\}$ cannot be sufficient.

The question is, at bottom, an instance of the measurement problem. But leaving deeper interpretative issues aside for the moment, an uncontroversial sufficient condition can be gleaned from a standard measurement scenario, which we take here as a motivating example. 

Suppose we have a particle in a double-well potential and an apparatus that interacts only briefly with the particle to perform a position measurement. The measurement concludes at time $t=0$ and records the particle as being in one of the two wells. Under the Schr\"odinger evolution for the total wave function, this results in an entangled superposition of particle (coordinate $y$) and apparatus (coordinate $x$) of the form 
\begin{equation}\label{measurement-state}\Psi(x,y, t) = \psi_1(x,t)\phi_1(y, t) + \psi_2(x, t)\phi_2(y,t),\end{equation}
where $\phi_1(y, t=0) \approx \phi_m(y), \; \phi_2(y, t=0) \approx \phi_a(y)$, and  $\psi_1, \psi_2$ are the corresponding ``pointer states'' of the apparatus, recording the outcome `$m$' or `$a$', respectively. In effect, the measurement prepares the particle in one of the localized states $\phi_m$ or $\phi_a$, which subsequently undergoes autonomous double-well dynamics governed by $H_y$. A key feature justifying this reading is decoherence induced by the pointer states---more precisely, that $\psi_1$ and $\psi_2$ remain localized in disjoint regions of configuration space, thus maintaining \emph{negligible overlap} in~$x$.

With this benchmark in place, we return to our original problem, which concerns not entanglement with a macroscopic apparatus or environment but entanglement between microscopic degrees of freedom. Nonetheless, we can formulate an analogous condition. We shall say that the $y$ system has a \emph{relative effective wave function} (REWF) if, and as long as, 
\begin{equation}\label{QEWF}
\Psi(x,y,t) \approx \sum\limits_{i=1}^{n} \psi_i(x,t)\phi_i(y,t), 
\end{equation}
where the components $\psi_i$, evolving under $H_x$, maintain \emph{persistently negligible overlap} in $x$-space. In this case, the $x$-components define (approximately) distinct non-interfering branches on which $\Psi$ is factorized, and we can expect---in analogy with the measurement situation---that the $y$-system is effectively described by one of the relative states $\phi_i$ evolving autonomously under $H_y$.

The Bohmian perspective, to be discussed in Section \ref{sec:BohmianCWF}, will make this argument precise. (In particular, to the extent that, in the measurement example, work is done by the idea that a macroscopic pointer has an \emph{actual} configuration, the same is also true of microscopic particle positions in Bohmian mechanics.) For now, we return our focus to the tunneling experiment of Sharoglazova et al.\ and explain why the REWF condition is realized in the propagative regime---where the naive tunneling time is correct---but not in the evanescent regime, where the naive tunneling time is false.

\subsection{The propagative regime}

The REWF condition \eqref{QEWF} is effectively realized in the propagative regime of the waveguide experiment. The wave packet in this regime is given by  
\be\label{prop}
\Psi(x,y,t) = \frac1{\sqrt{2\pi}}\int\limits_{k^\uparrow_c}^{\infty}\widehat{\psi}_\mathrm{in}(k) \, \psi_k(x,y) e^{-i\omega(k)t} \mathrm{d}k
\ee with $\widehat{\psi}_\mathrm{in}(k)$ as in \eqref{Gaussian} for  $k_0 > k^\uparrow_c=\hbar^{-1}\sqrt{2m(V_0+\hbar J_0)}$ and $\sigma_k \ll k_0 - k^\uparrow_c$, and with $\psi_k(x,y)$ as in \eqref{psikprop0} for $x>0$. In that region, \eqref{prop} thus takes the form 
\begin{equation}\begin{split}
&\Psi(x,y,t) \approx \frac1{\sqrt{2\pi}} \int\limits_{k^\uparrow_c}^\infty \widehat{\psi}_{\rm in}(k)\,\mathrm{t}(k)\,e^{-i\omega(k)t}
\frac{1}{\sqrt{2}} \Big(e^{ik_+ x} \phi_+(y)+e^{i k_- x}\phi_-(y)\Big)\mathrm{d}k\\
&=\frac1{\sqrt{2\pi}}\int\limits_{k^\uparrow_c}^\infty \widehat{\psi}_{\rm in}(k)\, \mathrm{t}(k)\,e^{i(\bar{k} x-\omega(k)t)}
\frac{1}{\sqrt{2}} \Big(e^{i\delta k \, x} \phi_+(y)+e^{-i \delta k \, x}\phi_-(y)\Big)\mathrm{d}k\\
&=\frac1{\sqrt{2\pi}}\int\limits_{k^\uparrow_c}^\infty  \widehat{\psi}_{\rm in}(k)\, \mathrm{t}(k)\,e^{i(\bar{k} x-\omega(k)t)}
\Big(\cos(\delta k \, x)\,\phi_m(y)+i \sin(\delta k \, x)\phi_a(y)\Big)\mathrm{d}k,
\end{split}\end{equation}
with $k_{\pm}=k_{\pm}(k)$ and $\mathrm{t}(k)$ as in \eqref{tmean}. Since $\widehat{\psi}_{\rm in}(k)$ is sharply peaked around $k_0$, we can approximate $\delta k(k)\approx \delta k(k_0)=: {\delta k_0}$ and $\bar{k}(k)\approx\bar{k}(k_0)=:\bar{k}_0$,  which leads to
\begin{equation}\label{pre-QEWF-prop}
\Psi(x,y,t)\approx \Psi_{\rm env}(x,t)\, \Big(\cos({\delta k_0}\,x)\,\phi_m(y)+i\sin({\delta k_0}\,x)\,\phi_a(y)\Big),
\end{equation}
with the longitudinal envelope
\begin{equation}
\Psi_{\rm env}(x,t):=\frac{1}{\sqrt{2\pi}} \int\limits_{k^\uparrow_c}^\infty \,\widehat{\psi}_{\rm in}(k)\, \mathrm{t}(k)\,e^{i\left(\bar{k}x-\omega(k)t\right)} \mathrm{d}k.
\end{equation}
This envelope propagates with a group velocity
\begin{equation}
v_g \approx \frac{\hbar \bar{k}_0}{m}
\end{equation}
and disperses only weakly. 

We can coarse-grain the total wave packet along small longitudinal intervals, decomposing it into subpackets 
\begin{equation}
\Psi(x, y,t)\approx \sum_i \Psi_i(x, y,t),
\end{equation}
where each $\Psi_i$ is an independent solution of the Schr\"odinger equation. The width $\Delta x$ of the subpackets (in $x$ dimension) is chosen such that 
\begin{equation}\label{widthineq}
{1}/{\bar{k}_0} \ll \Delta x \ll {1}/{{\delta k_0}}.
\end{equation}
The first inequality ensures that each subpacket contains many longitudinal
wavelengths and propagates approximately at the common group velocity $v_g$. The second inequality guarantees that the trigonometric terms [$\cos({\delta k_0}\, x)$ and $\sin({\delta k_0} \, x)$] in \eqref{pre-QEWF-prop} vary negligibly across the $x$-support of a given component, so that the $\Psi_i$ (approximately) factorize: 
\begin{equation}\label{subpacketfactorization}
\Psi(x, y,t)\approx \sum_i \Psi_i(x, y,t) \approx \sum_i \psi_i(x,t) \, \phi_i(y,t) \,. 
\end{equation}

Now, crucially, \eqref{widthineq} also implies that the longitudinal components $\psi_i(x,t)$ maintain negligible overlap over dynamically relevant timescales. Indeed, the time required for such a subpacket---moving at the group velocity $v_g$---to traverse a distance comparable to its own spatial width $\Delta x$ is
\begin{equation}
t_{\rm cross}\sim \frac{\Delta x}{v_g}
\approx \frac{m\,\Delta x}{\hbar \bar{k}_0}.
\end{equation}
Over this timescale, the broadening of the subpacket due to dispersion is (by the Heisenberg uncertainty relation) of order
\begin{equation}
\Delta x_{\rm disp}\sim \frac{\hbar t_{\rm cross}}{m\,\Delta x}
\approx \frac{1}{\bar{k}_0},
\end{equation}
which is small compared to $\Delta x$. Thus, on this timescale, the
$\psi_i$ remain well localized and only weakly overlapping. Over longer times, the subpacket decomposition might need to be iterated, but---by the argument just given---the REWF condition will be satisfied piecewise in time and ensure that the transverse dynamics are effectively autonomous double-well dynamics despite entanglement with the longitudinal degree of freedom.\footnote{More explicitly, denoting by $q_i(t)\approx q_i(0)+v_g t$ the center of $\psi_i(x,t)$ and using the identity $\bar{k}_0{\delta k_0}=\frac{m}{\hbar}J_0$, comparison between \eqref{pre-QEWF-prop} and \eqref{subpacketfactorization} yields \begin{equation}\phi_i(y, t) \approx
\bigl(\cos(J_0 t+c_i)\,\phi_m(y)
+i\sin(J_0 t+c_i)\,\phi_a(y)\bigr),
\end{equation} 
which are precisely the oscillatory solutions of the one-dimensional double well (cf. Eq. \eqref{1DdoubW}), differing by a constant phase offset $c_i:= {\delta k_0} \, q_i(0)$.}

\subsection{The evanescent regime}

By contrast, no such decomposition is available in the ring example introduced earlier. Looking back at \eqref{es1}, the standing waves $\sin(\bar{k}x)$ and $\cos(\bar{k}x)$ can be thought of as superpositions of counter-propagating plane waves $e^{i\bar{k}x}$ and $e^{-i\bar{k}x}$. If decomposed into smaller, localized subpackets, those would propagate in opposite directions and constantly collide, thus not maintaining negligible overlap.  

The REWF condition similarly fails in the evanescent regime of the waveguide experiment. Recall from \eqref{psi2Dint} that evanescent states in the classically forbidden region take the form  
\begin{equation}\begin{split}
&\Psi(x,y,t)\\
&\approx \frac{1}{\sqrt{2\pi}} \int\limits_0^{k^\downarrow_c} \widehat{\psi}_{\rm in}(k)  \gamma(k) e^{-i\omega(k)t} \frac{1}{\sqrt{2}} \left(e^{-\kappa_+(k) x} \phi_+(y)+e^{-\kappa_-(k) x}\phi_-(y)\right)\mathrm{d}k \\
&=  \frac{1}{\sqrt{2\pi}} \int\limits_0^{k^\downarrow_c}  \widehat{\psi}_{\rm in}(k) \gamma(k) e^{-i\omega(k)t} e^{- \bar{\kappa} x}\left(\cosh(\delta\kappa \, x) \phi_m(y) + \sinh(\delta\kappa \, x) \phi_a(y)\right)\mathrm{d}k.
\end{split}\end{equation} 
Here, the transverse modes are modulated by real exponential factors that overlap throughout the entire step region $x>0$, precluding any local factorization of the form \eqref{QEWF}. There is no longitudinal group velocity and no Rabi-style oscillations between the waveguides. The classically forbidden region rather acts as an energy-dependent filter:  as $x$ increases, the anti-symmetric double-well mode $\phi_-$ is attenuated more strongly than the symmetric ground state $\phi_+$ (since $E_+ < E_-$ implies $\kappa_+ < \kappa_-$). As a result, the relative population in the waveguides is controlled by the longitudinal distance from the step---at a spatial rate $\delta\kappa$ set by the differential attenuation---rather than by autonomous, time-dependent double-well dynamics.

\subsection{The Bohmian perspective: conditional wave functions}\label{sec:BohmianCWF}

Our analysis was guided by the following question. Given a decomposition
\be
\Psi(x,y,t) \approx \sum\limits_{i=1}^{n} \psi_i(x,t) \, \phi_i(y,t),
\ee
with $\psi_i$ and $\phi_i$ evolving unitarily under $H_x$ and $H_y$, respectively, under what conditions may the $y$-system be regarded as effectively described by one individual branch state $\phi_i$? The Bohmian perspective makes precise why the REWF condition provides a natural answer. 

In Bohmian mechanics, the $x$-coordinate has, at all times, an actual value $X(t)$. If and as long as the components $\psi_i(x, t)$ maintain negligible overlap in $x$-space, this $X(t)$ singles out one of the branches as actually guiding the system---and hence, in particular, the corresponding $\phi_i(y,t)$ as guiding the Bohmian dynamics of $Y(t)$. 

This is made precise by the concept of \emph{conditional wave functions} \cite{QE}. Given the full wave function $\Psi(x,y,t)$, the conditional wave function for the $y$-system is defined as
\begin{equation}\label{cwf}
\varphi(y,t)=\varphi^X(y,t) := \Psi(X(t), y, t).
\end{equation}
In terms of this wave function, the Bohmian guidance equation for $Y$ takes the usual form
\be
\dot{Y}(t) = \frac{\hbar}{m} \mathrm{Im} \frac{\nabla_y \varphi(y,t)}{\varphi(y,t)}\Bigl\lvert_{y = Y(t)}\,.
\ee
The conditional wave function is always well defined, but its time evolution depends, in general, explicitly on the evolution of $X(t)$. 

However, under the REWF condition, the conditional wave function behaves like a  wave function for the $y$-subsystem, evolving autonomously under $H_y$. If, at some time $t_0$, $X(t_0)$ lies in the ``support'' of, say, $\psi_1(x,t_0)$ (where $\psi_{i\neq 1}(X(t_0),t_0) \approx 0$), we find

\begin{equation}\begin{split}\label{effectivewf}
\varphi(y, t_0) &= \psi_1(X(t_0), t_0) \, \phi_1(y, t_0) + \sum\limits_{j \neq 1} \underbrace{\psi_j(X(t_0), t_0)}_{\approx 0} \, \phi_j(y,t_0) \\&\approx \psi_1(X(t_0), t_0) \, \phi_1(y, t_0),
\end{split}\end{equation}
and thus, after normalization, \begin{equation}\label{effectivewf2}\varphi(y, t_0) \approx \phi_1(y, t_0).\end{equation}
Two features ensure that \eqref{effectivewf} and hence \eqref{effectivewf2} continue to hold over relevant timescales, so that $\varphi(y, t)$ tracks the Schr\"odinger evolution of $\phi_1(y, t)$ under $H_y$.  
\begin{enumerate}
    \item The subpackets $\psi_i$ maintaining negligible overlap in $x$.
    \item The equivariance of the Bohmian particle dynamics, implying that (as long as the $\psi_i$ maintain negligible overlap) a typical $X(t)$ will continue to follow the same subpacket, i.e., stay in the $\psi_1$-branch. 
\end{enumerate} 
Hence, $X(t)$ selects the guiding branch but does not otherwise affect the dynamics of the conditional wave function for $y$. 

A measurement situation, as sketched around Eq.\ \eqref{measurement-state}, is the paradigmatic example. If, say, the actual pointer configuration $X(t)$ is in the support of the pointer state $\psi_1$---thus recording `$m$'---the conditional wave function of the particle in the double-well reduces to 
$\phi_1(t,y) = e^{-\frac{i}{\hbar}H_y t} \phi_m(y)$. Here, the relevant branching is provided by pointer states with macroscopically disjoint supports in configuration space. In such cases, the conditional wave function is also called an \emph{effective wave function}---which is the precise Bohmian analog of a pure state for the $y$-system \cite{QE}.

In the propagative regime of the experiment, we neither assume nor require macroscopic decoherence. As explained earlier, over short but characteristic time intervals, we obtain the decomposition \eqref{subpacketfactorization}, with the subpackets $\psi_i(t,x)$ maintaining negligible overlap. The actual longitudinal position $X(t)$ will typically follow one of those subpackets, and the conditional wave function for the transverse degree of freedom becomes the corresponding relative state, $\varphi(y,t) = \phi_i(y,t)$, which is effectively governed by autonomous double-well dynamics. 

More explicitly, one can readily see from \eqref{pre-QEWF-prop} that, along a typical Bohmian trajectory with $X(t) \approx v_g t$,
\begin{equation}\label{conditionalwf}
\varphi(y, t) \approx \Psi_{\rm env}\bigl(t,X(t)\bigr)\Big(\cos(J_0 t)\,\phi_m(y)+i\sin(J_0 t)\,\phi_a(y)\Big).
\end{equation}

\begin{figure}[ht]
\centering
  \includegraphics[width=\textwidth]{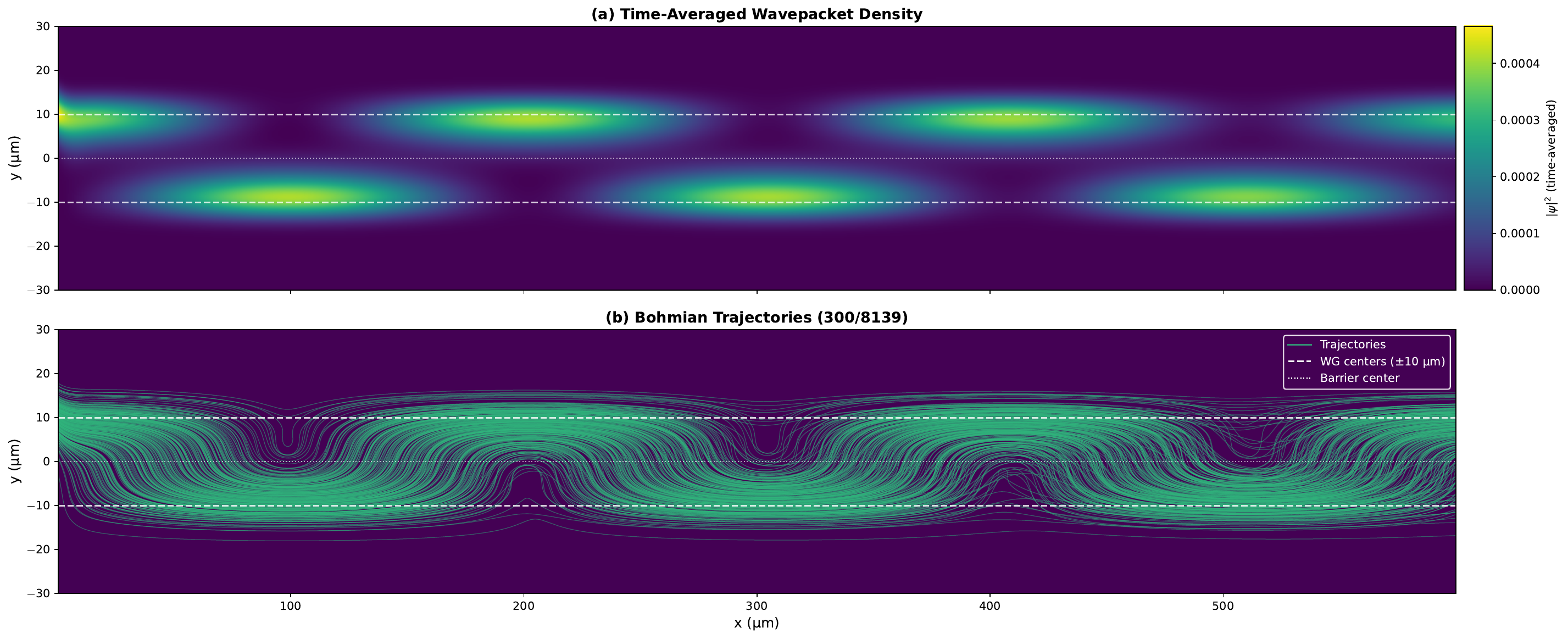}
  \caption{\label{Fig:propagative} Numerically obtained $|\Psi|^2$-density (top) and Bohmian trajectories (bottom) for $x>0$ in the propagative regime ($ E_{k_0}-V_0= +0.15\,\mathrm{meV}, \sigma_k = 2.5\times 10^{-3}\,\mu\mathrm{m}^{-1}$).}
\end{figure}

So, once again, we see that the conditional wave function of the $y$-system evolves according to $H_y$---with a double-well state undergoing coherent oscillations with angular frequency $J_0$. This is clearly reflected in the numerical plots of the oscillating wave packet and the corresponding Bohmian trajectories in Fig. \ref{Fig:propagative}. 

By contrast, in the evanescent regime, the conditional wave function $\varphi(y,t) \sim \cosh(\delta\kappa \, X(t)) \, \phi_m(y) + \sinh(\delta\kappa \, X(t)) \, \phi_a(y)$ retains a non-trivial dependence on $X(t)$. Consequently, it never evolves under autonomous double-well dynamics governed solely by $H_y$, but remains parametrically coupled to the Bohmian motion in the longitudinal direction.

We emphasize that, while Bohmian mechanics makes the point particularly transparent, the assumption of a tunneling time $\sim 1/J_0$ in the evanescent regime, and hence the speed estimate \eqref{longvelK2}, is already unjustified on the basis of Schr\"odinger dynamics alone. This stands in contrast to the suggestions of Sharoglazova et al. that the speed estimate \eqref{longvelK2} is a consequence of standard quantum mechanics.

\section{Resolution: Dwell Time and the B\"uttiker Formula}
\label{sec:dwell}

The dwell time argument presented in Section \ref{sec:dwelltime} seemed to corroborate the speed estimates inferred from the naive tunneling time. However, that argument suffers from a more basic flaw than the tunneling time reasoning. 

\subsection{Meaning of the formula}

The B\"uttiker dwell time formula represents an average over the full statistical ensemble---including the case in which particles  are reflected before entering the potential step and hence contribute \emph{zero} dwell time to this average. The desired speed estimate, on the other hand, concerns the speed \emph{inside the potential step}; for a speed estimate based on dwell time, the relevant dwell time would be an average \emph{conditional on penetrating the step} (and thus on spending a positive amount of time in the classically forbidden region). Since only a small fraction of the incident flux penetrates the step, \eqref{dwell1} grossly underestimates the dwell time conditional on penetration. Consequently, \eqref{vdwell} is a gross overestimate for particle speeds in the classically forbidden region.

While the tunneling time argument gave the right speed estimate in the propagative but not in the evanescent regime, the dwell time argument---in the form sketched in Section \ref{sec:dwelltime} and employed by Sharoglazova et al.---already fails in the propagative case. Indeed, consider first a free plane wave $\varphi_k=e^{ikx}$ and any bounded interval $[a,b]$; the  B\"uttiker dwell time formula \eqref{dwell1} for the mode $\varphi_k$ (which we repeat here for convenience)
\be\label{dwell3}
\tau_d(k) = (j_{\rm in}(k))^{-1} \int_a^b \mathrm{d}x\, |\varphi_k(x)|^2 \,,
\ee
then yields
\be
\tau_d(k) = \frac{ |b - a|}{j_{\rm in}(k)},
\ee
which agrees with the traversal time of a particle crossing the distance $|b-a|$ with constant velocity $v=j_{\rm in}(k)=\frac{\hbar k}{m}$.  However, inside a potential step, the propagative stationary scattering states are of the form $\varphi_k(x)=\mathrm{t}(k)e^{i\bar{k}x}$, with $k$ the wavenumber of the incident plane wave and $\bar{k}$ the wavenumber of the transmitted plane wave (see \eqref{propscatteringstates}).  The B\"uttiker dwell time, for $[a, b]$ inside the potential step, is thus 
\be
\tau_d(k) = \frac{ |\mathrm{t}(k)|^2 |b - a|}{j_{\rm in}(k)}\,.
\ee
This corresponds to the traversal time at velocity $v' = \hbar \bar{k}/m$---the correct velocity inside the potential step---multiplied by the transmission probability $T = |\mathrm{t}(k)|^2 \,{\bar{k}}/{k}$. 
In other words, the B\"uttiker dwell time is the \emph{average} dwell time over the \emph{whole ensemble}; it is defined relative to the total incoming flux $j_{\rm in}(k)$. To infer particle speeds \emph{conditional on entering the potential step}, it must be renormalized to the transmitted flux $j_{\rm tr}(k)= T j_{\rm in}(k)$, i.e., more generally, divided by the probability of entering the potential step.

In the following, we will provide a more precise statement of the B\"uttiker formula to clarify its status and meaning. Then we will show that the corrected dwell-time argument yields perfect agreement with the average Bohmian velocities in the evanescent region---turning what was supposed to be a challenge for Bohmian mechanics into an independent confirmation of its predictions.

\subsection{Origin of the formula}
\label{sec:buttikerformula}

The B\"uttiker formula is uncontroversial and is often taken as the very definition of dwell time for stationary scattering states. Its meaning, however, is clearer if one first emphasizes that it represents an \emph{average time}: more precisely, the mean value of the random variable \(T_{(a,b)}\), the total time spent in the region \((a,b)\), evaluated over an ensemble of particle trajectories whose positions are distributed according to \(|\psi(x,t)|^2\) at every time \(t\). This status and meaning are best clarified by the following mathematical fact.

\medskip

\noindent\textbf{Theorem (B\"uttiker formula).}\quad
Let $\psi(x,t)$ be a normalized solution of the Schr\"odinger equation 
with spectral representation

\be\label{wavepacketstationarystates}
\psi(x,t) =  \frac{1}{\sqrt{2\pi}} \int\limits_0^\infty \widehat{\psi}_{\rm in}(k)\, 
\varphi_k(x)\, e^{-i\omega(k)t}\, \mathrm{d}k,
\ee
where $\omega(k) = \hbar k^2/(2m)$, $\varphi_k(x)$ are generalized 
eigenfunctions of the Hamiltonian with unit incoming amplitude 
($\varphi_k(x) \approx e^{ikx}+\mathrm{r}(k) e^{-ikx}$ as $x \to -\infty$), and the spectral profile $\widehat{\psi}_{\rm in}(k)$ is supported on $(0,\infty)$. Let $(X(t))_{t \in \mathbb{R}}$ be an 
ensemble of trajectories with positions distributed according to 
$\rho(t,x) = |\psi(x,t)|^2$ at every time $t$. Then, for any 
$-\infty \leq a < b \leq +\infty$, the expected value of the time $T_{(a,b)}$ spent in 
$(a,b)$ is
\begin{align}
\tau_{(a,b)} &:= \mathbb{E}(T_{(a,b)}) = \mathbb{E} \int_{-\infty}^{+\infty} \mathrm{d}t\; 
\mathds{1}_{(a,b)}(X(t)) \\\label{def:dwell-time}
&= \int\limits_{-\infty}^{+\infty} \mathrm{d}t \int\limits_a^b \mathrm{d}x \; 
|\psi(x, t)|^2  \\\label{packetdwelltime}
&= \int_0^{\infty} {\mathrm{d}k}\, |\widehat{\psi}_{\rm in}(k)|^2\, 
\tau_d(k)
\end{align}
with $\tau_d(k)$ given by the B\"uttiker formula \eqref{dwell3}.

\medskip

\noindent A more general result will be stated and proven in a separate paper. The identity between \eqref{def:dwell-time} and \eqref{packetdwelltime} is known (see, e.g., \cite{Hauge.etal1987, LeavensAers1989}) and follows most easily from the Plancherel theorem applied to the Fourier transform in the time variable, together with the change of variables $\omega = \hbar k^2/2m$. 

Three observations further clarify the scope and interpretation of the B\"uttiker dwell time:

\begin{enumerate}
\item The formula assumes that positions at any time $t$ are 
Born-distributed, with $\psi$ normalized to unity. This applies to the complete ensemble of incoming trajectories, including 
those that never enter $(a,b)$. To obtain the average dwell 
time conditional on entering $(a,b)$, one must divide by the 
probability of entry.
\item It is important to stress that no further restriction on the ensemble of trajectories is required. In particular, the trajectories need not be Bohmian. The only assumption is that, at each time \(t\), their positions are distributed according to \(|\psi(x,t)|^2\). This condition is compatible with a vast class of possible stochastic or deterministic processes, and does not single out any specific dynamical law for the trajectories.

\item For a wave packet with spectral profile sharply peaked around $k_0 >0$, \eqref{packetdwelltime} yields
\be
\int_0^{\infty} \mathrm{d}k\, |\widehat{\psi}_{\rm in}(k)|^2\, 
\tau_d(k) \approx \tau_d(k_0) \int_0^{\infty} \mathrm{d}k\, |\widehat{\psi}_{\rm in}(k)|^2 = \tau_d(k_0).
\ee
Thus, the B\"uttiker dwell time \eqref{dwell3} is 
the monochromatic limit of the average dwell time for (normalized) incoming wave packets. Note, however, that for a stationary state $\psi = \varphi_{k}\, e^{-i\omega(k)t}$, which is not normalized (and not normalizable), the time integral in \eqref{def:dwell-time}
diverges.
 
\end{enumerate}

\subsection{The dwell time argument corrected}

Returning to the experiment of Sharoglazova et al., we recall from \eqref{dwell2} that the B\"uttiker formula \eqref{dwell3} for a particle in a 1D  setup with a potential step (i.e., with $(a,b) = (0, \infty)$ and $\varphi_{k}(x)=\frac{2ik}{ik-\kappa}e^{-\kappa x}$) yields 
\begin{equation}\label{Buttiker2}
\tau_d(k)
= \frac{2 m k}{\hbar \kappa (k^2+\kappa^2)}
\end{equation}
for the dwell time in the classically forbidden region $x>0$.

Interpreted correctly, this corresponds (in the monochromatic limit) to the expected dwell time conditional on entering the classically forbidden region, multiplied by the probability of doing so.

To estimate the probability of penetrating the potential step, we recall from \eqref{timereversal} that the time evolution of the wave packet in the evanescent region is symmetric about the central time $t=0$, when the amplitude in the forbidden region goes from increasing to decreasing, corresponding to a reversal of the probability flux. The probability of entering the forbidden region should thus correspond to the total probability in the forbidden region at $t=0$. While the precise value will be calculated below, a simple order-of-magnitude estimate is given by the ratio of the evanescent decay length $\lambda=\frac{1}{2\kappa_{0}}$ (roughly the width of the wave packet in the forbidden region) to the width $\sigma_x = {\sigma_k}^{-1}$ of the incoming wave packet, yielding  \begin{equation}P_{\rm pen} \sim\frac{\sigma_k}{\kappa_0}.\end{equation} 
The dwell time \emph{conditional on entering the forbidden region} is thus \begin{equation}\label{taue}
\tau_{\rm pen} := \frac{\tau_d}{P_{\rm pen}} \sim \frac{2mk_{0}}{\hbar \sigma_k (k_{0}^2 + \kappa_{0}^2)}.    
\end{equation} Dividing twice  the typical penetration depth $\lambda=\frac{1}{2\kappa_{0}}$ by this time (in line with the dwell time argument in Section \ref{sec:dwelltime}), the corrected speed estimate is
\begin{equation}\label{dta-corrected}
v_x \sim \frac{2\lambda}{\tau_{\rm pen}} \sim \frac{\hbar\sigma_k}{m}\left(\frac{1+\beta_{0}^2}{2\beta_{0}}\right), 
\end{equation}
with $\beta_{0}= k_{0} / \kappa_{0}$. 
Comparing \eqref{dta-corrected} to \eqref{vdwell}, we see that the factor $\kappa$ in the numerator is replaced by the spectral width $\sigma_k$. The speed thus vanishes in the monochromatic limit---consistent with the fact that a stationary evanescent wave carries no probability current.   

Using \eqref{psi2trev2}, we can obtain a more exact expression for $P_{\rm pen}$ in the small $\sigma_k$ limit: 
\begin{align}\label{eq:Pmax}
P_{\rm pen} = \int_0^{\infty} |\psi(x,0)|^2\, \mathrm{d}x \approx \frac{\gamma_0^2\,\sigma_k}{\sqrt{2\pi}\kappa_0}.
\end{align}
From \eqref{rtkappasub}, we get $\gamma_0^2 = \frac{4k_0^2}{(k_0^2+\kappa_0^2)}$ and thus 
\begin{align}\label{Ppen}
P_{\rm pen}  \approx \frac{2 \sqrt{2}\,  k_0^2}{\sqrt{\pi}\, (k_0^2+\kappa_0^2)}\,\frac{\sigma_k}{\kappa_0}. 
\end{align}
Using this more exact expression for $P_{\rm pen}$, we obtain, instead of \eqref{taue}, that

\begin{equation}\label{taupen}
\tau_{\rm pen} = \frac{\tau_d(k_0)}{P_{\rm pen}} \approx \frac{2 m k_0}{\hbar \kappa_0 (k_0^2+\kappa_0^2)} \frac{\sqrt{\pi} (k_0^2+\kappa_0^2) \kappa_0}{2\sqrt{2}\, k_0^2 \sigma_k} = \sqrt{\frac{\pi}{2}}\frac{m}{\hbar k_0 \sigma_k},
\end{equation}
in complete agreement with \eqref{sojourn}, the mean dwell time of Bohmian trajectories that enter the classically forbidden region. 
The associated improved velocity estimate is
\begin{equation}\label{idta-corrected}
v_x \sim \frac{2\lambda}{\tau_{\rm pen}} \approx \sqrt{\frac{2}\pi}\frac{\hbar k_0\sigma_k}{m\kappa_0} =  \sqrt{\frac{2}\pi}\frac{\hbar\sigma_k}{m} \beta_{0},
\end{equation}
which agrees, apart from  a  factor of $\pi/4$, with the average Bohmian velocity \eqref{vB} in the forbidden region. (The discrepancy is due to the fact that the expected value of a ratio need not agree with the ratio of the expected values.)

We note that the corrected speed \eqref{idta-corrected} scales with $\beta =  {k}/{\kappa}$, in contrast to the evanescent speed \eqref{longvelK2} proposed by Sharoglazova et al., which is proportional to $\kappa$. The correction thus resolves their paradoxical conclusion that ``the lower the energy of the particles, the faster they move within the step potential'' \cite[p. 69]{Sh25}.

The derivation of \eqref{taupen} and \eqref{idta-corrected} did not hinge on Bohmian mechanics. However, as in our discussion of tunneling times, the Bohmian perspective makes the analysis more precise. First, the B\"uttiker formula, as stated in Section \ref{sec:buttikerformula}, has a clear physical content in terms of particle trajectories with Born-distributed positions---precisely the setting provided by Bohmian mechanics (cf. \cite{leavens1990}).

Second, the parabolic form of the Bohmian trajectories justifies the heuristic argument for the penetration probability $P_{\rm pen}$. Since all trajectories entering the forbidden region turn around at the same time $t=0$, the probability of entering the forbidden region is the probability of being there at that time, i.e., \eqref{Ppen}. 

Third, the conditional dwell time $\tau_{\rm pen} = \tau_\mathrm{B}$ can be computed directly from the Bohmian trajectories as in \eqref{sojourn}, providing an independent route to the B\"uttiker dwell time: $\tau_{\rm pen}\cdot P_{\rm pen} = \tau_d(k_0)$.  Note that we have a penetration probability of order $\sigma_k$ and an average dwell time of Bohmian trajectories---conditional on entering that region---of order $\sigma_k^{-1}$. Their product thus survives in the limit $\sigma_k \to 0$, even though each factor individually degenerates ($P_{\rm pen} \to 0$ and $\tau_\mathrm{B} \to +\infty$, respectively). 

In any case---whether derived from the general B\"uttiker formula, or directly from the residence times of Bohmian trajectories---$\tau_d(k_0)$ is precisely the Bohmian prediction, in the narrow-band limit, for the unconditional average dwell time of trajectories, i.e., the average over the complete ensemble of particles, including those that are reflected before entering the potential step at $x=0$.

\begin{figure}[h!]
\centering
  \includegraphics[width=\textwidth]{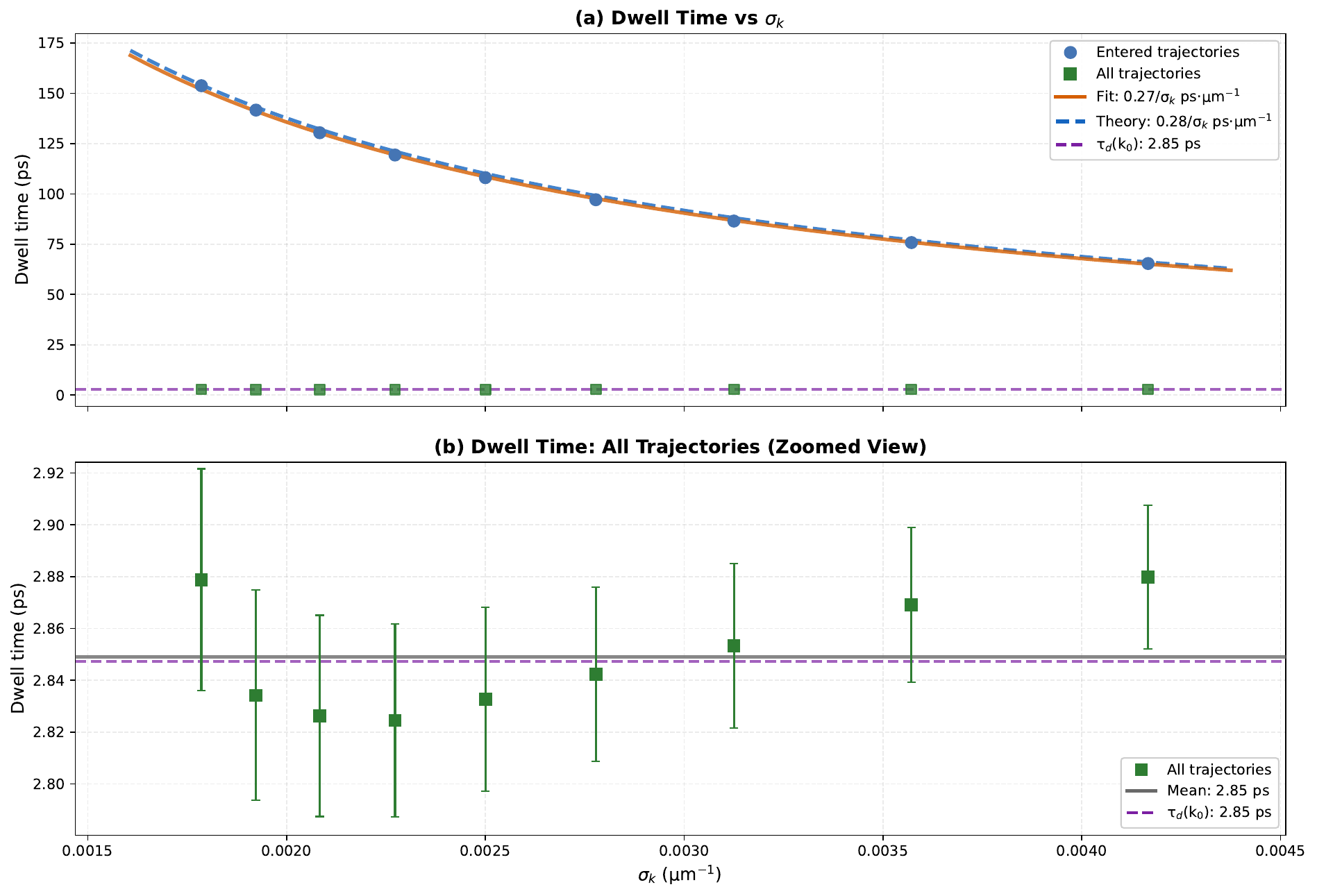}

  \caption{  \label{fig:Bohmiandwelltimes} Dwell time versus initial momentum spread $\sigma_k$ at $ E_{k_0} -V_0 = -0.09\,\mathrm{meV}$. 
(a) Average dwell times of trajectories entering the forbidden region (blue circles) exhibit $1/\sigma_k$ scaling. 
The empirical fit (solid orange, $0.271\,\mathrm{ps{\cdot}\mu m^{-1}}/\sigma_k$) agrees 
with the theoretical estimate \eqref{sojourn} (dashed blue, 
$0.275\,\mathrm{ps{\cdot}\mu m^{-1}}/\sigma_k$) to within $\sim 1.5\%$. Green squares: unconditional dwell times (all trajectories). 
Purple dashed line: 1D B\"uttiker dwell time $\tau_d(k_0)$ at central momentum. 
Error bars ($\pm 1$ SE) smaller than markers. (b) Zoomed view of all trajectories 
($n = 300{,}000$ per point) shows the unconditional dwell times clustering around 
$\tau_d(k_0) = 2.85\,\mathrm{ps}$. (Note that the probed range of $\sigma_k$ lies significantly above the corresponding value $\sigma_k \approx 10^{-5}\,\mu\mathrm{m}^{-1}$ in the experiment of \cite{Sh25}, cf. footnote \ref{footnote6}.)}
\end{figure}

Numerical simulations show excellent agreement with these theoretical predictions, see Fig. \ref{fig:Bohmiandwelltimes}.

\section{Conclusion}
\label{sec:conclusions}

We have examined two velocity estimates along the lines of Sharoglazova et al.\ \cite{Sh25}---one based on the tunneling time between parallel waveguides, and the other on the B\"uttiker dwell time---both of which conflict with the Bohmian predictions for particle motion in the classically forbidden region.

The tunneling time procedure yields the (uncontroversially correct) group velocity in the propagative regime, but its applicability does not extend to the evanescent regime. As explained in Section \ref{sec:tunnelingtime}, entanglement between the longitudinal and transverse degrees of freedom disrupts the double-well dynamics, and the naive tunneling time (set by the waveguide coupling $J_0$) no longer applies. The conflict is therefore not between Bohmian mechanics and experiment---or between Bohmian mechanics and orthodox quantum mechanics---but between Bohmian mechanics and a flawed heuristic underlying the inference from waveguide populations to particle speeds. 

The original dwell time argument already goes wrong in the propagative regime since the B\"uttiker formula represents an average over all incoming particles, only some of which actually enter the potential step. Once corrected---by conditioning on the particles entering the potential step---it yields the standard group velocity in the propagative regime. Applied in the evanescent regime, the same corrected procedure yields precise quantitative agreement with the Bohmian prediction in the narrow-band limit. 

What appeared as a two-pronged challenge to Bohmian mechanics thus resolves, on closer analysis, into a confirmation of the theory: the procedure that yields the correct answer in the propagative regime and extends to the evanescent regime recovers exactly the Bohmian prediction.

Notably, the flaws in the original speed estimates can already be identified from the viewpoint of standard quantum mechanics. The Bohmian analysis, however, makes them particularly transparent. For the tunneling time argument, the Bohmian concept of conditional wave functions makes precise why the naive tunneling time is appropriate in the propagative regime but not in the evanescent regime. For the dwell time argument, the Bohmian trajectories give precise meaning to the B\"uttiker formula and allow us to derive the correct dwell times from first principles. 

The deeper moral may be that if one wants to know how fast particles are moving in a classically forbidden region, it is good to consult a theory that actually describes particles in motion.

\paragraph{Funding.---} 
The work of C.B. and D.L. was supported by the Israel Science Foundation (grant no.\ 1597/23).

\paragraph{Code Availability.---} 
The numerical code used to generate the plots in this work is available from D.L. upon reasonable request.

\paragraph{Competing interests.---}  The authors declare no competing interests. 

\paragraph{Author contributions.---} All authors contributed equally to the research and writing of this manuscript.

\appendix

\section{Approximation of the Bohmian velocity in the 2D  model \label{app:meanvel}}

Starting from the narrow-band approximation \eqref{psi2D_shited} of the two-dimensional evanescent wave packet, we want to estimate the longitudinal Bohmian velocity $v_x^\Psi(x,y,t)
=
\frac{\hbar}{m}\Im\!\left(\frac{\partial_x\Psi}{\Psi}\right)$ in the weak-splitting regime $\delta\kappa \ll \bar{\kappa}$. 

Recall that $\beta_\pm = {k_0}/{\kappa_\pm}$. We introduce
\be
\bar\beta:=\frac{\beta_++\beta_-}{2},
\qquad
\delta\beta:=\frac{\beta_+-\beta_-}{2},
\ee
so that $\beta_\pm=\bar\beta\pm\delta\beta$ and
$
\frac{\delta\beta}{\bar\beta}=\frac{\delta\kappa}{\bar{\kappa}}\ll1.
$

With
\begin{equation}
\psi_0(x,t):=
\gamma_0\left(\frac{\sigma_k^2}{2\pi}\right)^{1/4} e^{-\bar\kappa x}e^{-i\omega_0 t}
\exp\!\left[
{\sigma_k^2}\,
\Bigl(
(\bar\beta x-i\omega_0't)^2+(\delta\beta x)^2
\Bigr)
\right],
\end{equation}
the longitudinal factors (pertaining to the even and odd transverse modes, respectively) can be written as
\begin{equation}
\psi_\pm(x,t)=\psi_0(x,t)\,e^{\pm q(x,t)},
\qquad
q(x,t):=
\delta\kappa\,x
+2\sigma_k^2\delta\beta\,x\bigl(\bar\beta x-i\omega_0't\bigr).
\end{equation}
Hence
\begin{align}
\Psi(x,y,t)&=\psi_0(x,t)\,\Xi(x,y,t),\\
\Xi(x,y,t)&:=\frac{1}{\sqrt2}\Bigl(e^{q}\phi_+(y)+e^{-q}\phi_-(y)\Bigr),
\end{align}
and therefore
\be
v_x^\Psi
=
\frac{\hbar}{m}\Im\!\left(\frac{\partial_x\psi_0}{\psi_0}\right)
+
\frac{\hbar}{m}\Im\!\left(\frac{\partial_x\Xi}{\Xi}\right).
\ee
The first term yields 
\begin{equation}
\frac{\hbar}{m}\Im\!\left(\frac{\partial_x\psi_0}{\psi_0}\right)
=
-\frac{\hbar}{m}2\sigma_k^2\bar\beta\,\omega_0't.
\label{eq:leadpsi0}
\end{equation}
To estimate the second term, we write \(q=u-iv\) with:
\be
u=
\delta\kappa\,x+2\sigma_k^2\delta\beta \bar\beta x^2,
\qquad
v=2\sigma_k^2\delta\beta\,x\,\omega_0't.
\ee
Since \(\phi_\pm\) are real-valued, setting
\be
A:=e^{u}\phi_+(y),\qquad B:=e^{-u}\phi_-(y),
\ee
we obtain
\be
\frac{\partial_x\Xi}{\Xi}
=
(\partial_x q)\,
\frac{e^{q}\phi_+-e^{-q}\phi_-}{e^{q}\phi_++e^{-q}\phi_-}
=
(\partial_x q)\,
\frac{(A-B)\cos v-i(A+B)\sin v}{(A+B)\cos v-i(A-B)\sin v}.
\ee
On any compact set away from the nodes of \(\Psi\), the last quotient has real part \(O(1)\) and imaginary part \(O(v)\). Moreover,
\be
\partial_x q
=
u'-iv'
\quad \text{with} \quad
u'=
\delta\kappa+ 4\sigma_k^2\delta\beta \bar\beta x,
\qquad
v'=2\sigma_k^2\delta\beta\,\omega_0't.
\ee
Therefore, 
\be
\Im\!\left(\frac{\partial_x\Xi}{\Xi}\right)
=
\Im\!\bigl((u'-iv')\cdot O(1+iv)\bigr)
=
O(v')+O(u'v),
\ee
and both these terms are (at most) $O(\delta\beta\,\omega_0') \sigma_k^2 t$, hence 
\begin{equation}
\Im\!\left(\frac{\partial_x\Xi}{\Xi}\right)
=
O\!\bigl(\delta\beta\,\omega_0'\bigr) \sigma_k^2 t.
\label{eq:Xicorr}
\end{equation}

\noindent Combining \eqref{eq:leadpsi0} and \eqref{eq:Xicorr} (and recalling that ${\delta\beta}/{\bar\beta}={\delta\kappa}/{\bar{\kappa}}$), we get
\be
v_x^\Psi(x,y,t)
=
-\frac{\hbar}{m}2\sigma_k^2\bar\beta\,\omega_0't
\left(
1+O\!\left(\frac{\delta\kappa}{\bar{\kappa}}\right)
\right),
\ee
which is a more precise statement of \eqref{meanvel}.

\end{document}